\pdfoutput=1  

\documentclass[11pt]{article}

\usepackage[margin=1in]{geometry}
\providecommand{\TitleFont}{\LARGE\bfseries}  

\usepackage[T1]{fontenc}       
\usepackage{lmodern}           
\usepackage{amsmath}
\usepackage{amssymb}
\usepackage{graphicx} 
\usepackage{subcaption} 
\usepackage{xcolor} 
\usepackage{cite}
\usepackage{url}
\usepackage{hyperref}
\hypersetup{
    colorlinks=true,
    linkcolor=blue,
    citecolor=blue,
    urlcolor=black
}

\newcommand\blfootnote[1]{%
  \begingroup
  \renewcommand\thefootnote{}\footnote{#1}%
  \addtocounter{footnote}{-1}%
  \endgroup
}

\title{\TitleFont Narrow-Shell Stochasticity in Source--Sink Models of the Low Earth Orbit Environment}

\author{Jaewon Choi \\ Dept. of Industrial and Management Engineering, Myongji University\\ \href{mailto:j1choi@mju.ac.kr}{j1choi@mju.ac.kr} \and Souvik Dhara \\ H. Milton Stewart School of Industrial \& Systems Engineering, Georgia Institute of Technology \\
\href{mailto:souvik.dhara@isye.gatech.edu}{souvik.dhara@isye.gatech.edu} \and Harsha Honnappa \\ Edwardson School of Industrial Engineering, Purdue University \\ \href{mailto:honnappa@purdue.edu}{honnappa@purdue.edu}}

\date{}

\begin{document} 

\maketitle

\begin{abstract}
\normalsize
Deterministic source--sink models are widely used to assess the
long-term evolution, capacity, and sustainability of the low Earth orbit (LEO) environment. 
These models propagate shell-averaged populations through ordinary differential equations (ODEs), implicitly relying on individual collision, disposal, and decay events to average out within sufficiently large altitude shells. 
As constellation traffic is increasingly organized into kilometer- and sub-kilometer-scale shells, this averaging assumption becomes increasingly strained. 
We formulate the multi-shell, multi-species LEO environment as a Markov jump process and, from this stochastic formulation, recover the conventional source--sink ODE as a large-volume limit and derive a stochastic differential equation (SDE) approximation whose fluctuations scale as $V^{-1/2}$ in the per-shell volume $V$. 
The diffusion approximation is validated against an exact discrete-event simulation of the underlying jump process. 
Sweeping $V$ at fixed spatial density over the $450\,\mathrm{km}$--$800\,\mathrm{km}$ band, we find that the two descriptions agree at shell volumes comparable to those used in established source--sink models but diverge as shells narrow. 
At the finest shell volume considered, the mean debris population reaches roughly $4.5$ times the ODE prediction, with several realizations undergoing runaway growth absent from the ODE trajectory. 
The departure from the ODE trajectory is driven by the nonlinear collision terms, through which population variance and covariance raise expected collision rates, generating further debris and reinforcing the collision--debris feedback. 
At sufficiently fine shell volumes, narrow-shell stochasticity therefore both widens the distribution of possible outcomes and alters the expected trajectory. 
Because the stochastic model shares its parameterization with the deterministic one, it provides a scale-consistent extension of existing source--sink models for evaluating shell configuration, collision risk, and long-term LEO sustainability. \blfootnote{\emph{Acknowledgments:} JC acknowledges the Purdue IE Summer Internship Program 2024, where the major part of this work was done. HH acknowledges Abhi Deshmukh IE Frontiers Grant, from the Edwardson School of Industrial Engineering.}
\end{abstract}

\section{Introduction}
\label{sec:intro}

The population of Anthropogenic space objects (ASOs), including active satellites, derelicts and debris, in low-Earth orbit (LEO) evolves temporally through collisions, drag, and launch processes. 
Source--sink models, often referred to as ``particle-in-a-box'' frameworks, have become an instrumental tool for the long-term assessment of the ASO population and to evaluate sustainability of the LEO environment. In a nutshell, these models partition LEO into orbital shells and use systems of ODEs to describe the evolution of the shell-averaged densities of ASO populations across these shells. This approach reduces the problem of tracking tens of thousands of individual objects to a substantially lower-dimensional dynamical system that can be efficiently propagated over long time horizons to assess capacity limits, collision-risk thresholds, and post-mission disposal (PMD) targets~\cite{d2024carrying, d2023novel, long2020impacts}. The equilibrium points of these ODE systems also provide a natural characterization of sustainable states of the LEO environment~\cite{d2024carrying}.

The accuracy of ODE-based predictions rests on an averaging assumption. When orbital shells are sufficiently wide, numerous deterministic and stochastic events occur within a given time window, allowing their individual effects to average out and the aggregate dynamics to be well approximated by ODEs. Existing source--sink models are built on this premise and typically employ shells that are $10$--$50\mathrm{km}$ wide. However, the wide-shell assumption is increasingly at odds with the relevant scales in current and emerging LEO operations and policy. The radial extent assigned to individual constellations is contracting toward the kilometer scale \cite{lifson2025}.  Capacity studies increasingly divide LEO into thin orbital shells. By carefully arranging traffic within these shells, adjacent shells can be separated by only a few kilometers, and in some analyses by less than one kilometer \cite{arnas2021,lifson2024}. Regulatory practice is also shifting toward narrower shells: the Federal Communications Commission (FCC) has authorized SpaceX shells separated by only $5\mathrm{km}$~\cite{fcc2022gen2}. Operational studies similarly consider fine separation scales, with Iridium-related analyses estimating required centerline spacings of $1.8$--$6.5\mathrm{km}$~\cite{lifson2025}. At such scales, ODE-based modeling is limited in its applicability to LEO policy decisions, where its underlying deterministic averaging assumption fails to hold.
\begin{quote}
    In this paper we demonstrate that, at narrower shell widths, individual stochastic events no longer average out, leading to long-term predictions that differ drastically from those of ODE-based models.
\end{quote}

\paragraph*{Methods and Contributions.}
This demonstration first requires a formalization of the dynamics of ASOs in the LEO environment as a Markov jump process (MJP). The formulation is inspired by compartmental models of epidemic spread \cite{Draief_Massoulie_2009}.
The model tracks three species, namely, active satellites ($S$), derelicts ($D$), and debris ($N$), in each orbital shell. Each encounter causes the corresponding populations to increase or decrease, with encounter rates between different species governed by a kinetic collision kernel. Encounters may result in lethal, disabling, or no-effect outcomes, captured through physically interpretable coefficients. Debris yields follow the NASA Standard Breakup Model~\cite{Krisko2011Proper}, while atmospheric drag induces a Markovian downward flux between adjacent shells. Launches are modeled as independent Poisson processes, and post-mission disposal (PMD) acts on active satellites with a configurable success probability.

To connect the MJP with existing source--sink models, we first show that, as the shell volume $V$ grows, the species densities converge to the solution of the ODE system underlying the MOCAT-3 model~\cite{d2024carrying,d2023novel}. This establishes that the MJP can be viewed as a microscopic stochastic model whose macroscopic limit corresponds to MOCAT-3. Next, we derive a finer-scale approximation of the MJP that captures the $O(V^{-1/2})$ finite-volume fluctuations neglected by the ODE limit. More precisely, using the framework of functional central limit theorems~\cite{Draief_Massoulie_2009,ethier2009markov}, we show that the suitably rescaled fluctuations of the species densities converge to a system of stochastic differential equations (SDEs). The resulting It\^o system has independent Brownian motion drivers associated with each collision mechanism, drag flux, launch process, and PMD event. Since the ODE and SDE approximations are both derived from the same underlying MJP, their predictions can be compared directly, allowing us to quantify the impact of the finer-scale stochastic fluctuations captured by the SDEs.

\begin{figure}[t]
  \centering
  \includegraphics[width=0.5\textwidth]{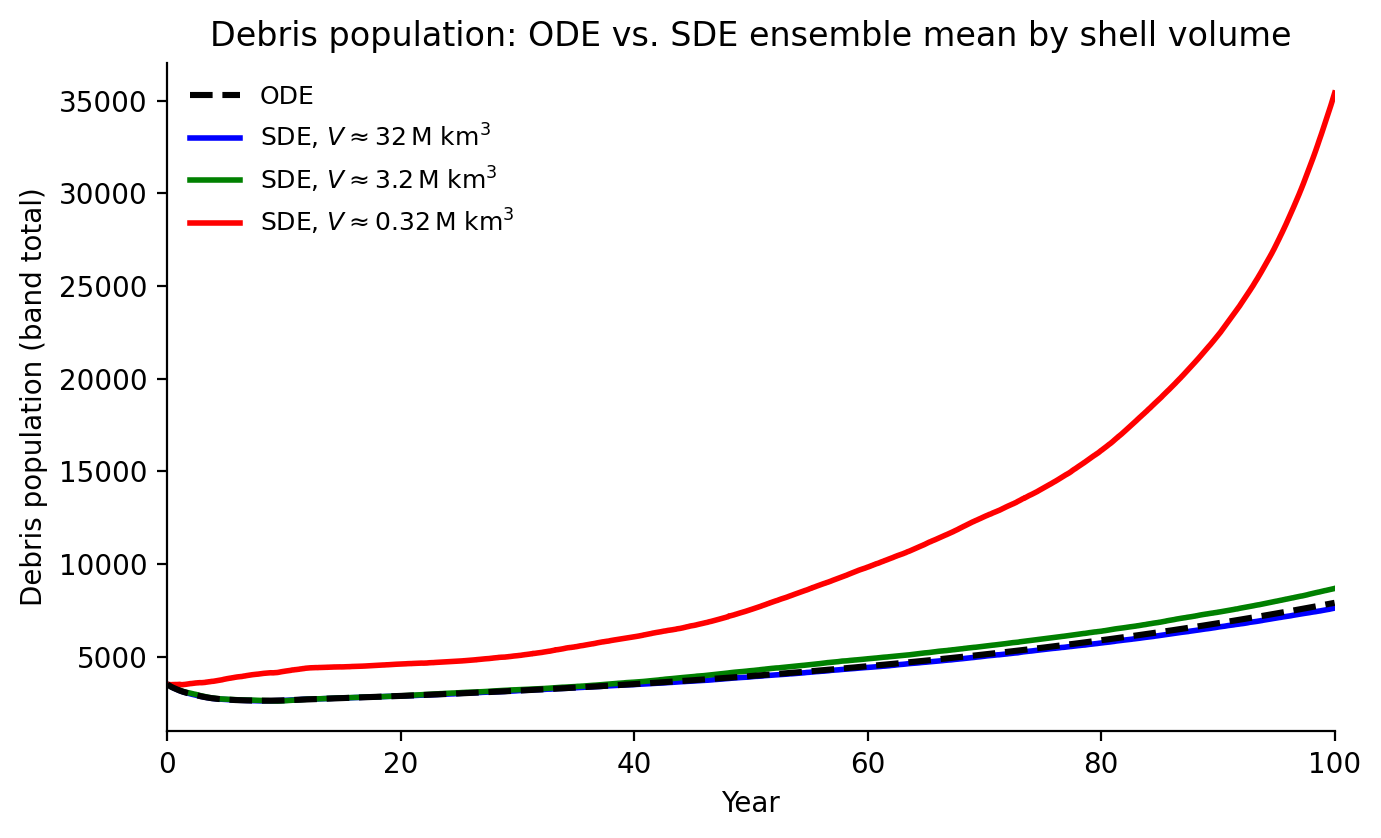}
  \caption{Ensemble-mean debris population over the $100$-year horizon at
  three shell volumes, against the ODE trajectory. At $0.32\,\mathrm{M\,km^3}$ the mean departs to several times the ODE
  value. Detailed in Section~\ref{sec:results}.}
  \label{fig:intro_debris}
\end{figure}

The forecasts from the ODE and SDE are compared over a 100-year horizon for decreasing shell volumes of three orders of magnitude, namely $V = 32, 3.2,$ and $0.32$ million cubic-kilometers ($\mathrm{~M~km}^3$), respectively. 
At $V = 32\mathrm{~M~km}^3$, the ODE approximation is highly accurate, and the ODE and SDE predictions are nearly indistinguishable. At $V = 3.2\mathrm{~M~km}^3$, although the mean SDE prediction remains close to the ODE prediction, the SDE ensemble exhibits substantial variability, with standard deviation being roughly twice its mean. This suggests that, while the ODE may accurately capture the average behavior, its deterministic prediction may not reliably characterize individual realizations. Finally, at $V = 0.32\mathrm{~M~km}^3$, the mean SDE prediction at the end of the horizon is approximately $4.5$ times the corresponding ODE value. Nearly half of the realizations finish above three times the ODE value, and a few grow without bound. See Fig.~\ref{fig:intro_debris} which shows the ensemble-mean debris trajectory at each shell volume against the ODE.
Thus, at this shell volume, SDEs provide a first-order shift in predictions rather than a second-order correction to a
deterministic mean. This is one of the key takeaways of this paper. 
We acknowledge that $V = 0.32\mathrm{~M~km}^3$ is an extreme case, but the objective is to stress test the predictions.
In Section~\ref{sec:discussion}, we discuss how the gap between the two predictions need not be confined to the variance, and how collisions increase the expected collision rate, which in turn generates further fragments. 

The findings show that a shell deemed safe by the ODE framework may still carry a non-negligible probability of runaway growth once finer scale approximations are taken into account. By linking operational decisions, such as launch scheduling and avoidance efficacy, to environment-wide risk, the framework provides a tool for stress-testing capacity and policy conclusions against realistic stochastic dynamics. These findings also bear on ongoing work on shell design and slotting~\cite{lifson2023space}, where tighter packing may increase intrinsic capacity while potentially heightening sensitivity to rare events.

The rest of the paper is organized as follows. Section~\ref{sec:related} situates the
paper among existing modeling schemes. Section~\ref{sec:model} introduces
the compartment model and its source--sink structure.
Section~\ref{sec:mjp} formulates the Markov jump process and derives the
deterministic ODE limit per species. Section~\ref{sec:sde} develops the
stochastic differential equation approximation and its drag and launch
terms. Section~\ref{sec:sims} describes the simulation setup---initial
population, launch-rate estimation, and the shell-volume scaling
design---and presents the deterministic baseline against the stochastic
ensemble. Section~\ref{sec:results} reports the three findings and
their check against an exact discrete-event sampler.
Section~\ref{sec:discussion} interprets them, locates the shell volume
at which the deterministic description ceases to be adequate, and
expresses the resulting gap in policy units as the launch-cadence or
disposal-reliability adjustment needed to recover the ODE outcome.
Section~\ref{sec:conclusion} concludes and outlines remaining work on
constellation design and slotting.

\section{Related Work}
\label{sec:related}

Modeling the ASO population in LEO spans object-by-object propagation to aggregate compartmental flows. Established work clusters into categories of deterministic source--sink evolution, object-level propagation, and stochastic source--sink models.

Deterministic source--sink models trace to the collision-cascade analysis of Kessler and Cour-Palais~\cite{kessler1978collision} and extend to the JASON study of large constellations, featuring satellite, derelict, and debris species~\cite{long2020impacts}. 
The MOCAT family formalized the LEO environment as a multi-shell, multi-species source--sink system. The three-species variant MOCAT-3 and its four-species extension MOCAT-4N became the reference for capacity and stability analyses computed shell by shell ~\cite{d2023novel,d2024carrying}.
Throughout, populations are propagated as smooth, shell-averaged flows over altitude bins tens of kilometers thick. Analytic studies on deterministic source--sink evolution show that long-term evolution behavior depends on launch cadence and satellite placement~\cite{somma2019,pasiecznik2022}. 
Shell governance and coordination have been assessed based on the deterministic evolution models~\cite{lifson2023space}.
 
Object-level models propagate individual objects under orbital mechanics and test each candidate collision through explicit pairwise probability calculations.
NASA's LEGEND~\cite{liou2004} propagates each object's orbit from known launch and breakup records to simulate the historical population.
The Monte Carlo branch of MOCAT (MOCAT-MC)~\cite{jang2024mocatmc} propagates future scenarios, ensembling independent runs to report uncertainty.
 
Stochastic source--sink frameworks recast the shell populations as random processes, sampling the distribution of outcomes.
Fanto et al. treat launch, disposal, decay, and collision as random events, assessing how launch rate and PMD compliance drive the risk within an altitude shell~\cite{fanto2023}.

This work extends the deterministic approach into the stochastic source--sink category.
The stochastic dynamics are derived analytically from the same equations that yield the deterministic regime.
We then expose how the fluctuations scale across shell volumes. At smaller shells, they approximate the individual-event stochasticity that object-level models capture only by propagating objects one at a time.
The analysis marks where the deterministic description stays reliable and where it breaks down.

\section{Compartment Model}
\label{sec:model}

We model the LEO environment as a compartment model. 
ASO populations are partitioned into three qualitatively distinct species and the altitude range is partitioned into a stack of concentric shells of identical volume. 
Within each shell, the populations of each species evolve through collisions, atmospheric drag, launches,
and PMD. 
This section describes the species, the transition dynamics, and the source--sink structure.

\subsection{Species and Shell Geometry}

Following the source--sink convention \cite{long2020impacts, d2024carrying,
d2023novel}, we distinguish three species:
\begin{itemize}
  \item \textbf{Satellites} ($S$): active, functioning, maneuverable objects;
  \item \textbf{Derelicts} ($D$): non-functional intact objects such as spent payloads and rocket bodies that no longer maneuver;
  \item \textbf{Debris} ($N$): the collection of fragments produced by collisions and breakups.
\end{itemize}
We assume that only satellites are launched and disposed of, only satellites can avoid collisions, and derelicts and debris are removed solely by atmospheric drag.

The altitude band of interest is divided into $N_{\mathrm{sh}}$ shells,
each a thin space between concentric spheres. 
All shells are configured with the same volume $V$, with corresponding radial thickness $d$ which acts as the main control parameter for the study while keeping the spatial density of objects fixed. 
Each shell carries a state vector
\begin{equation}
  P_h(t) = \bigl(S_h(t),\, D_h(t),\, N_h(t)\bigr),
  \qquad h = 1, \dots, N_{\mathrm{sh}},
\end{equation}
collecting the satellite, derelict, and debris counts in shell $h$ at time $t$.

\subsection{Transition Dynamics}

\begin{figure}[t]
  \centering
  \includegraphics[width=.7\textwidth]{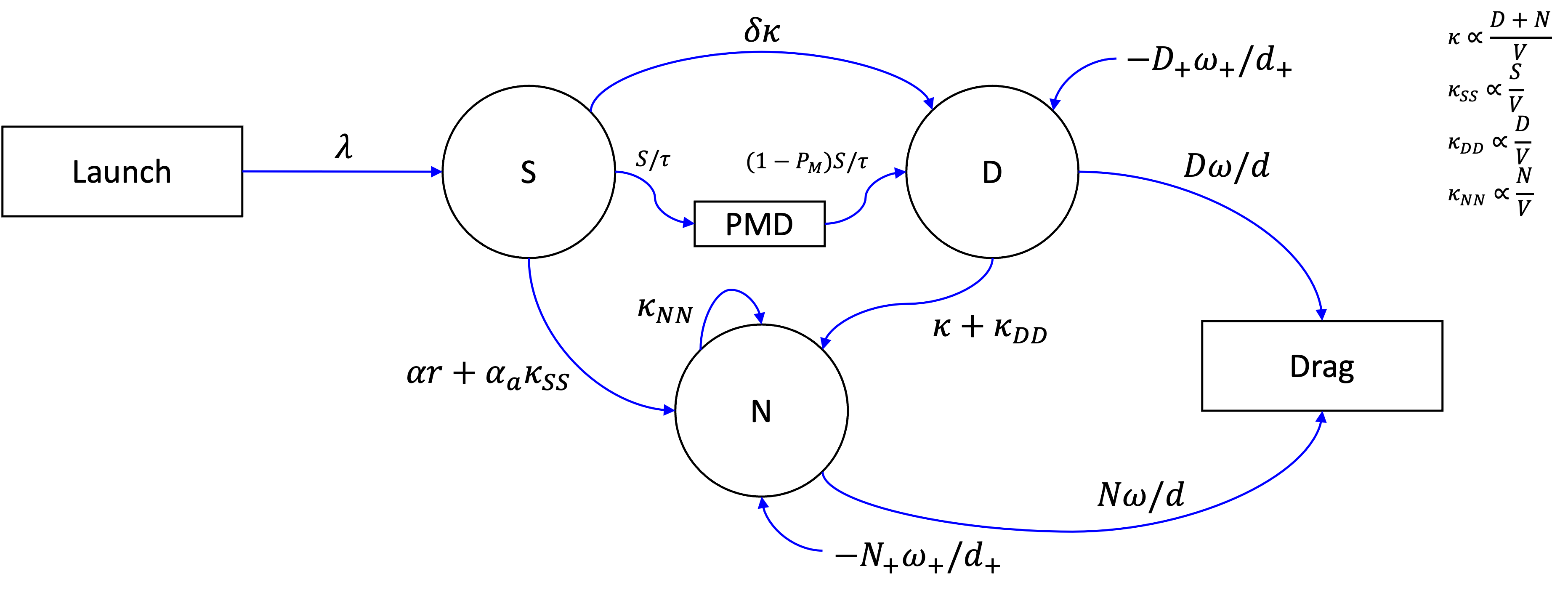}
  \caption{Compartment model of the LEO system. Solid arrows denote transitions among satellites ($S$), derelicts ($D$), and debris ($N$), each labeled with its schematic rate; $\kappa$ denotes a collision rate, subscripts marking the colliding species.}
  \label{fig:LEO_compartment}
\end{figure}

Five processes govern the evolution of $P_h(t)$, summarized in
Fig.~\ref{fig:LEO_compartment}.

\paragraph{Collisions.} We model collisions based on the kinetic theory of gases, as adopted in previous works \cite{d2024carrying, somma2017statistical, Trisolini2021Analysis, letizia2016}. Pairwise encounters between any two species occur at a rate set by a kinetic collision kernel
$\sigma_{ij} = (r_i + r_j)^2$, and the mean relative speed $v$. $\sigma_{ij}$ is a squared impact parameter defined as the squared sum of the radii $r_i$, $r_j$ of the colliding objects $i$ and $j$. Each encounter resolves into one of three outcomes: \emph{lethal} collisions, which destroy the participating intact objects and inject fragments according to the breakup model; \emph{disabling} collisions, which degrade an active satellite into a derelict without full fragmentation; and \emph{no-effect} encounters, which an active satellite avoids. 
Two coefficients carry this resolution: $\alpha$, the
fraction of satellite collisions with debris or derelicts that the satellite
fails to avoid, and $\delta$, the ratio of disabling to lethal outcomes.
Satellite--satellite collisions are governed by a separate avoidance factor
$\alpha_a$, reflecting that two active objects can coordinate to avoid one
another.

\paragraph{Debris generation.} Lethal and disabling collisions produce fragments
following the NASA Standard Breakup Model \cite{Krisko2011Proper}, under which the number of fragments
scales with the masses and the characteristic length of the colliding objects.
Catastrophic (fragmentation) and non-catastrophic (cratering) collisions are
distinguished by their fragment yields, so the debris source term depends on
which species pair collides and on the collision outcome.

\paragraph{Atmospheric drag.} Derelicts and debris are subject to atmospheric drag, which lowers their semi-major axis and removes them from their shell at a per-object rate $\omega/d$, where $\omega$ is the drag-induced rate of altitude loss.
Objects leaving shell $h$ enter the adjacent lower shell $h-1$, making drag the sole mechanism that couples shells. Satellites are assumed to maintain their altitude and are not subject to passive drag decay.
We assume that the derelict and debris populations of the uppermost shell are fixed at their initial values, which supplies a steady inflow to the lower shells.

\paragraph{Launch.} New satellites enter the system through launches, modeled as
an exogenous Poisson process with shell-specific rate $\lambda_h$ derived from
historical launch records and an empirical altitude distribution, which is described in detail in Section~\ref{sec:sims}.

\paragraph{Post-mission disposal (PMD).} When satellites reach the end of the operational lifetime $\tau$, they are removed through PMD with success probability $P_M$. 
A failed disposal routes into the derelict population.

\subsection{Governing ODEs}
\label{sec:governing_odes}
 
The evolution of ASO population in each shell decomposes into launch, PMD transitions, collision gains and losses, and drag-driven inter-shell flux:
\begin{equation}
  \dot{P}_h(t)
  = \dot{\Lambda}_h(t)
  + \dot{C}_{\mathrm{PMD},h}(t)
  + \dot{C}_h(t)
  + \dot{R}_h(t),
  \label{eq:sourcesink}
\end{equation}
where $\dot{\Lambda}_h$ is new launches, $\dot{C}_{\mathrm{PMD},h}$
collects the disposal and disposal-failure transitions, $\dot{C}_h$ captures the collision-induced gains and losses across species, and $\dot{R}_h$ is the net drag flux between shell $h$ and its neighbors. 
Launches act as a source for the active population; collisions move mass from intact species into debris and derelicts; and drag is a sink for each shell.
 
Equation~\eqref{eq:sourcesink} is the standard source--sink skeleton shared with established models such as MOCAT-3 \cite{d2023novel}. 
The governing equations are formulated in terms of the densities of each species within a shell $x_S$, $x_D$, and $x_N$. 
The evolution of satellites is described as follows. One satellite is lost to lethal and disabling collisions with debris and derelicts (combined factor $\delta+\alpha$), and two at a time to satellite--satellite collisions. 
PMD removes a satellite at rate $x_S/\tau$, and launches replenish a satellite at rate $\lambda$.
\begin{equation}
  \frac{dx_S}{dt} = \lambda
    - (\delta+\alpha)\,\sigma_{NS} v\, x_N x_S
    - (\delta+\alpha)\,\sigma_{DS} v\, x_D x_S
    - 2\,\alpha_a\,\sigma_{SS} v\, x_S^2
    - \frac{x_S}{\tau}.
  \label{eq:ode_S}
\end{equation}
 
For derelict dynamics, a derelict is lost to collisions with debris, and two are lost from derelict--derelict collisions. 
A derelict is gained when a collision disables a satellite, or when a satellite fails post-mission disposal. 
Drag removes derelicts to the shell below and replenishes them from the shell above.
\begin{equation}
  \begin{aligned}
    \frac{dx_D}{dt} =\ &
      -\,\sigma_{DN} v\, x_D x_N
      -\,\alpha\,\sigma_{DS} v\, x_D x_S
      +\,\delta\,\sigma_{DS} v\, x_S x_D
      +\,\delta\,\sigma_{NS} v\, x_S x_N \\
    & -\,2\,\sigma_{DD} v\, x_D^2
      +\,(1-P_M)\,\tfrac{x_S}{\tau}
      -\,\tfrac{\omega}{d} x_D
      +\,\tfrac{\omega_+}{d_+} x_{+D}.
  \end{aligned}
  \label{eq:ode_D}
\end{equation}
 
Debris is generated by collision and is removed by drag to the shell below and replenished from the shell above. 
The fragment yield per event is based on the NASA Standard Breakup Model \cite{Krisko2011Proper}: $n_{f,c} = 0.1\, L_C^{-1.71} (M_i + M_j)^{0.75}$ for catastrophic collision ($S \times S, S \times D, D \times D$) and $n_{f,nc} = 0.1\, L_C^{-1.71} (M_p\, v_{\mathrm{imp}}^2)^{0.75}$ for non-catastrophic collision ($S \times N, D \times N, N \times N$). 
$M_i$ and $M_j$ are the masses of the colliding species, $L_C$ is the minimum fragment characteristic length, $M_p = \min(M_i, M_j)$, and $v_{\mathrm{imp}}$ is the impact velocity.
\begin{equation}
  \begin{aligned}
    \frac{dx_N}{dt} =\ &
      n_{f,nc}\bigl(\sigma_{DN} v\, x_D x_N + \alpha\,\sigma_{NS} v\, x_N x_S + \sigma_{NN} v\, x_N^2\bigr) \\
    & + n_{f,c}\bigl(\alpha\,\sigma_{DS} v\, x_D x_S + \alpha_a\,\sigma_{SS} v\, x_S^2 + \sigma_{DD} v\, x_D^2\bigr)
      - \tfrac{\omega}{d} x_N + \tfrac{\omega_+}{d_+} x_{+N}.
  \end{aligned}
  \label{eq:ode_N}
\end{equation}
 
The breakup model is evaluated on the masses of the colliding objects. The collision probability is determined by the radii of the interacting objects and the fragment yield is determined by mass of the colliding objects. Derelicts and debris are distinguished by a representative mass and radius, so that the impact parameters $\sigma_{ij}$ and the fragment yields $n_{f,c}$, $n_{f,nc}$ are constants attached to each collision channel.
 
\section{Markov Jump Process and Deterministic Limit}
\label{sec:mjp}

The governing ODEs of Section~\ref{sec:model} treat each shell population as a smooth density evolving in continuous time. 
In reality, the population changes discretely every time an event occurs at random moments. 
This section describes this discrete process and derives the ODEs as its limit. 
Section~\ref{sec:mjp_def} formulates the population in a shell as a Markov jump process (MJP), whose transitions are the launch, collision, disposal, and drag events of Section~\ref{sec:model}. 
Section~\ref{sec:mjp_limit} then shows that, as the shell volume $V$ grows, the presented MJP recovers the governing ODEs.

\subsection{Markov Jump Process}
\label{sec:mjp_def}
 
The number of objects in an orbital shell changes in discrete steps. A launch adds one satellite, a collision removes one or two intact objects and injects a burst of fragments, and drag carries one object into the shell below. A Markov jump process tracks these integer counts together with the random times at which they change, and its state summarizes the shell population at each instant.

Let $\hat{X}_V(t)$ denote a Markov jump process of the population counts in a shell of volume $V$. 
The state of the process at time $t$ is the population vector $\hat{X}_V(t) = (\hat{X}_{V,S}(t), \hat{X}_{V,D}(t), \hat{X}_{V,N}(t))$, recording the number of satellites, derelicts, and debris, and the process moves in continuous time as these counts change.
Dividing by the shell volume gives the random density $\hat{x}_V(t) = \hat{X}_V(t)/V$, with components $\hat{x}_{V,Q}(t) = \hat{X}_{V,Q}(t)/V$ for species $Q \in \{S, D, N\}$.

Each type of event that can occur is called a channel, indexed by $c$. 
When a channel-$c$ event occurs, the state changes by a fixed jump size $\ell_c$. 
The intensity $\beta_c(x)$ of channel $c$ is the per-volume event rate of that channel, a function of the current densities $x$. 
Multiplying by the shell volume $V$ gives the rate $V\beta_c(x)$, the expected number of channel-$c$ events in the shell per unit of time. The channels, with their jump sizes and intensities, are enumerated in Table~\ref{tab:allchannels}.

 
\begin{table}[t]
  \centering
  \caption{Full jump-channel enumeration. Per-shell rates are $V\beta_c$. Fragment yields $n_{f,c}$ and $n_{f,nc}$ follow
the NASA Standard Breakup Model as in Section~\ref{sec:mjp}.}
  \label{tab:allchannels}
  \begin{tabular}{lcc}
    \hline
    Channel & $(\Delta\hat{X}_{V,S}, \Delta\hat{X}_{V,D}, \Delta\hat{X}_{V,N})$ & $\beta_c(x)$ \\
    \hline
    Satellite--debris lethal      & $(-1,\ 0,\ +n_{f,nc})$ & $\alpha\,\sigma_{NS}\, v\, x_S x_N$ \\
    Satellite--debris disabling   & $(-1,\ +1,\ 0)$        & $\delta\,\sigma_{NS}\, v\, x_S x_N$ \\
    Satellite--derelict lethal    & $(-1,\ -1,\ +n_{f,c})$ & $\alpha\,\sigma_{DS}\, v\, x_S x_D$ \\
    Satellite--derelict disabling & $(-1,\ +1,\ 0)$        & $\delta\,\sigma_{DS}\, v\, x_S x_D$ \\
    Satellite--satellite lethal   & $(-2,\ 0,\ +n_{f,c})$  & $\alpha_a\,\sigma_{SS}\, v\, x_S^2$ \\
    Derelict--debris collision    & $(0,\ -1,\ +n_{f,nc})$ & $\sigma_{DN}\, v\, x_D x_N$ \\
    Derelict--derelict collision  & $(0,\ -2,\ +n_{f,c})$  & $\sigma_{DD}\, v\, x_D^2$ \\
    Debris--debris collision      & $(0,\ 0,\ +n_{f,nc})$  & $\sigma_{NN}\, v\, x_N^2$ \\
    Launch                        & $(+1,\ 0,\ 0)$         & $\lambda$ \\
    PMD success                   & $(-1,\ 0,\ 0)$         & $P_M\, x_S/\tau$ \\
    PMD failure                   & $(-1,\ +1,\ 0)$        & $(1-P_M)\, x_S/\tau$ \\
    Drag, derelict out            & $(0,\ -1,\ 0)$         & $(\omega/d)\, x_D$ \\
    Drag, derelict in             & $(0,\ +1,\ 0)$         & $(\omega_+/d_+)\, x_{+D}$ \\
    Drag, debris out              & $(0,\ 0,\ -1)$         & $(\omega/d)\, x_N$ \\
    Drag, debris in               & $(0,\ 0,\ +1)$         & $(\omega_+/d_+)\, x_{+N}$ \\
    \hline
  \end{tabular}
\end{table}

The number of events that occur over a given period is random, even when their average rate is known. 
A Poisson process $Y$ provides a mathematical framework for such random event counts. 
We consider a unit-rate Poisson process, for which one event occurs on average per unit time. 
Thus, $Y(\theta)$ denotes the random number of events occurring by time $\theta$, with an expected count of $\theta$.

In our model, however, the event rate depends on the current state of the
system. For each channel $c$, let $Y_c$ denote an independent unit-rate
Poisson process. The per-shell event rate of channel $c$ at time $u$ is
$V\beta_c(\hat{x}_V(u))$, the intensity defined above scaled by the shell
volume \cite{wallace2012}. 
The integrated intensity over the interval $[0,t]$ is given by
\begin{equation}
  \theta_c^V(t) = \int_0^t V\beta_c\!\left(\hat{x}_V(u)\right) du,
  \label{eq:accumulated_rate}
\end{equation}
and $Y_c$ composed with $\theta_c^V(t)$ is a Cox process. In particular, $Y_c (\theta_c^V(t))$ gives the number of channel-$c$ events by time $t$.
 
Multiplying each event count by its jump size $\ell_c$ and summing over channels gives the population,
\begin{equation}
  \hat{X}_{V,D}(t) = \underbrace{\hat{X}_{V,D}(0)}_{\text{initial count}}
    + \sum_{c} \underbrace{\ell_c}_{\substack{\text{jump} \\ \text{size}}} \times 
      \underbrace{Y_{c}\!\left(\theta_c^V(t)\right)}_{\text{event count for channel } c \text{ by time } t}.
  \label{eq:jumpprocess}
\end{equation}
Equation~\eqref{eq:jumpprocess} states that the population starts from its initial value and adds the accumulated effect of every event.

\subsection{ODEs as Large-Volume Limits}
\label{sec:mjp_limit}
 
Taking $V \to \infty$, the jump process, rescaled by $V$, converges to the ODEs of Section~\ref{sec:model}. 
A short calculation for the derelict population $\hat{X}_{V,D}$ shows how the ODEs emerge; the satellite and debris counts follow by the identical construction.
 
Consider the events that remove one derelict over a short time interval $h$, namely a collision with debris, a collision with a satellite that the satellite fails to avoid, or removal by drag.
Each of the $D$ derelicts is treated as an independent Bernoulli trial, undergoing one of these events with a small probability proportional to $h$ or remaining unchanged otherwise. 
The probability that exactly one derelict is removed is then
\begin{equation}
  \begin{aligned}
    &P\bigl(\hat{X}_{V,D}(t+h) = D-1 \,\big|\, \hat{X}_V(t) = (S,D,N)\bigr) \\
    &\qquad = \underbrace{\binom{D}{1}\Bigl(\sigma_{DN} v\, \tfrac{N}{V} h\Bigr)
        \Bigl(1 - \sigma_{DN} v\, \tfrac{N}{V} h\Bigr)^{D-1}}_{\text{collision with a debris object}}
      + \underbrace{\binom{D}{1}\Bigl(\alpha\,\sigma_{DS} v\, \tfrac{S}{V} h\Bigr)
        \Bigl(1 - \alpha\,\sigma_{DS} v\, \tfrac{S}{V} h\Bigr)^{D-1}}_{\text{collision with an unavoided satellite}}
      + \underbrace{\tfrac{\omega}{d} D\, h}_{\text{removal by drag}} + o(h).
  \end{aligned}
  \label{eq:D_minus1}
\end{equation}
Each binomial factor $\binom{D}{1}(\cdot)(1-\cdot)^{D-1}$ is the probability that exactly one of the $D$ derelicts triggers an event.
 
The remaining derelict transitions are constructed identically. 
Disabling collisions, PMD failure, and drag from above each contribute a gain term. 
A derelict--derelict collision removes two objects at once, so it carries jump size $\ell_c = -2$. 
Refer to Appendix~\ref{app:channels} for full derelict dynamics.
 
We now turn the probability into an expectation. 
Over the interval $h$, each channel $c$ changes the derelict count by $\ell_c$ with probability $V\beta_c(\hat{x}_V)h + o(h)$.
Multiplying each jump size by its probability and summing over channels gives the expected net change,
\begin{equation}
  \mathbb{E}\bigl[\hat{X}_{V,D}(t+h) - \hat{X}_{V,D}(t) \,\big|\, \hat{X}_V(t)\bigr]
    = V h \sum_{c} \ell_c\, \beta_c(\hat{x}_V) + o(h).
  \label{eq:expected_increment}
\end{equation}
Dividing by $V$ converts counts to densities, and dividing by $h$ and
letting $h \to 0$ gives a derivative. 
The derelict density evolves according to the net drift
\begin{equation}
  \frac{dx_D}{dt} = \sum_{c} \ell_c\, \beta_c(x) =: F_D(x),
  \label{eq:drift_def}
\end{equation}
in which each channel contributes its intensity weighted by the size of the jump it produces.
 
The Kurtz--Ethier theorem \cite{ethier2009markov,Draief_Massoulie_2009} states that as $V \to \infty$, the random density $\hat{x}_V(t)$ converges uniformly on compact time intervals to a deterministic density $x(t)$, the solution of \eqref{eq:drift_def}. The formal statement is given in Appendix~\ref{app:channels}. Collecting the three species, the density process $\hat{x}_V(t)$ is approximated in this limit by the deterministic system
\begin{equation}
  \frac{dx}{dt} = F(x), \qquad F(x) = \bigl(F_S(x),\, F_D(x),\, F_N(x)\bigr), \qquad F_Q(x) = \sum_{c} \ell_c\, \beta_{c,Q}(x),
  \label{eq:ode_limit}
\end{equation}
whose components are the governing ODEs~\eqref{eq:ode_S}--\eqref{eq:ode_N} of Section~\ref{sec:model}. Equivalently, the total population count is approximated by the deterministic density scaled by the shell volume,
\begin{equation}
  \hat{X}_V(t) \approx V\, x(t),
  \label{eq:ode_population}
\end{equation}
with the approximation exact in the limit $V \to \infty$. 


\section{Stochastic Differential Equation Approximation}
\label{sec:sde}

The deterministic equation of Section~\ref{sec:mjp} propagates the population as a mean flow. 
This section augments that equation with a fluctuation term derived from the same Markov jump process. 
The two descriptions then share a common parameterization and differ only in the event-level randomness. 
Whether the mean flow is an accurate summary of the dynamics depends on the per-shell volume $V$. 
In a large-volume shell, the events occurring within the shell are numerous enough to average out over the horizon of interest. 
The mean flow is then a representative summary of the dynamics. 
Contemporary constellations, however, are filed into far smaller-volume shells. 
Each shell then carries proportionally fewer objects and hosts proportionally fewer events. This motivates the finer scale approximations, driven by the SDEs described below. 
 
\subsection{Magnitude of the Fluctuations}
\label{sec:sde_scaling}
 
Let $\beta_c$ denote the intensity per unit volume of the collision channel $c$ defined in Section~\ref{sec:mjp}. 
The number of channel-$c$ events over an interval $\Delta t$ is Poisson-distributed with mean $V\!\beta_c\,\Delta t$, and therefore deviates from that mean by an amount of order $\sqrt{V\!\beta_c\,\Delta t}$. 
This absolute deviation grows with the shell volume. The size of the deviation relative to the mean it perturbs governs the accuracy of the mean-flow description.
\begin{equation}
  \frac{\sqrt{V\!\beta_c\,\Delta t}}{V\!\beta_c\,\Delta t}
  = \bigl(V\!\beta_c\,\Delta t\bigr)^{-1/2}
  \;\sim\; V^{-1/2},
  \label{eq:counting}
\end{equation}
This deviation from the mean scales as the inverse square root of the number of events the shell hosts. 
For example, reducing a shell volume in half, and the number of events it contains in proportion, raises the relative fluctuation by a factor of $\sqrt{2}$, exposing the fluctuation within the shell.
 
\subsection{The Diffusion Approximation}
\label{sec:sde_model}
 
Retaining this fluctuation alongside the mean flow yields the SDE approximation $Z_V$ to the random density $\hat{x}_V$,
\begin{equation}
  Z_{V,Q}(t) = x_Q(0)
    + \int_0^t F_Q(Z_V(u))\, du
    + \frac{1}{\sqrt{V}} \sum_{c} \ell_c
      \int_0^t \beta_{c,Q}^{1/2}(Z_V(u))\, dW_{c,Q}(u),
  \label{eq:Z}
\end{equation}
for each species $Q$, where the $W_{c,Q}$ are independent Brownian motions, one per channel. Their form follows from the deviation introduced in Section~\ref{sec:sde_scaling}: accumulated over many collision events, the functional central limit theorem for density-dependent Markov chains~\cite{ethier2009markov} shows that this departure behaves as a Brownian motion of amplitude $V^{-1/2}$, which is precisely the scaling carried by the diffusion term above. The formal construction, including the drag and launch terms, is given in Appendix~\ref{app:sde}. 

The drift $F_Q$ is exactly the deterministic source--sink equation of Section~\ref{sec:mjp}, and letting $V \to \infty$ recovers the ODE term by term. The diffusion term is the new content. It assigns to each channel $c$ an independent noise contribution whose amplitude is the square root of that channel's own event rate, weighted by its jump size $\ell_c$ and scaled by the $V^{-1/2}$ factor of \eqref{eq:counting}. 
This construction holds identically for every channel, including drag and launch. 

Multiplying \eqref{eq:Z} by the shell volume gives the population estimate
\begin{equation}
  \hat{X}_{V,Q}(t) \approx V\, Z_{V,Q}(t) = V\, x_Q(t) + \sqrt{V}\, \sum_{c} \ell_c
    \int_0^t \beta_{c,Q}^{1/2}(Z_V(u))\, dW_{c,Q}(u)
  \label{eq:Z_population}
\end{equation}
The population count is the deterministic term $V x_Q(t)$ of \eqref{eq:ode_population} augmented with a fluctuation of order $\sqrt{V}$.

The square-root amplitude carries a physical interpretation. The
fluctuation is not an external disturbance imposed on the dynamics but a property of the dynamics themselves, arising because collisions occur as discrete events rather than as a continuous flow. Its size is consequently determined by the current state of the shell. A shell with low collision rates carries correspondingly small fluctuations, whereas a shell that evolves into a debris-rich, collision-heavy state acquires larger ones, and these in turn feed the collisions that generate them. The diffusion term thus participates in the debris--collision feedback, which distinguishes \eqref{eq:Z} from the deterministic
equation.
 
This feedback also displaces the ensemble mean. Although the diffusion term has zero mean, the collision intensities are quadratic in the populations, so a fluctuation raises the collision rate by more than an equal fluctuation of the opposite sign lowers it. Averaged over realizations, a shell therefore undergoes more collisions, and produces more fragments, than one held at the mean population, and the imbalance grows with the fluctuation amplitude, hence as shells narrow. The ensemble mean consequently lies systematically above the deterministic trajectory, by a margin that widens as $V$ decreases. This behavior is measured directly through the shell-volume sweep of Section~\ref{sec:sims}.
 
\section{Simulation Setup}
\label{sec:sims}

We simulate the multi-shell LEO system over a 100-year horizon from a
2025 baseline, using the ODE and SDE formulations of
Sections~\ref{sec:mjp} and~\ref{sec:sde}. The simulated band spans
$450\,\mathrm{km}$--$800\,\mathrm{km}$, which carries roughly $80\%$ of the ASO
population and contains the altitudes into which current and planned
constellations are filed; it is therefore both the most populated part
of LEO and the part where shell widths are contracting fastest. This
section describes the initial population, the data-driven launch rate,
the shell-volume scaling design that is the empirical core of the paper,
and the numerical solvers.

\subsection{Initial Population}

\begin{figure}[t]
  \centering
  \includegraphics[width=.5\textwidth]{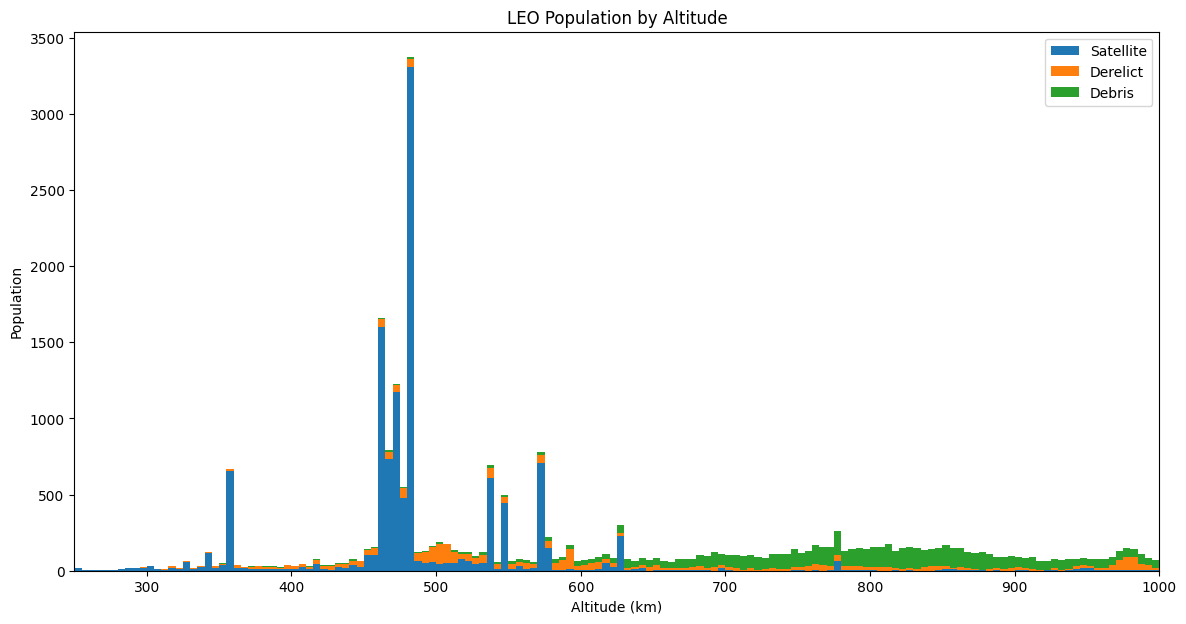}
  \caption{Initial population of satellites, derelicts, and debris across LEO
  altitudes, derived from McDowell satellite statistics \cite{mcdowell2025_active_sats}.}
  \label{fig:init_pop}
\end{figure}

The initial orbital population is drawn from satellite statistics compiled by
McDowell \cite{mcdowell2025_active_sats}, which provide the counts of active
satellites, derelicts, and debris across discrete altitude ranges as of 2025. The population counts are fitted with a continuous altitude-dependent distribution, as illustrated in Fig.~\ref{fig:init_pop}. Each altitude shell in the simulation is then initialized with species-specific populations proportional to this fitted distribution, so that the per-shell density is preserved across shell configurations. 

\subsection{Launch Rate}
\label{sec:launch_rate}

\begin{figure}[t]
  \centering
  \includegraphics[width=.5\textwidth]{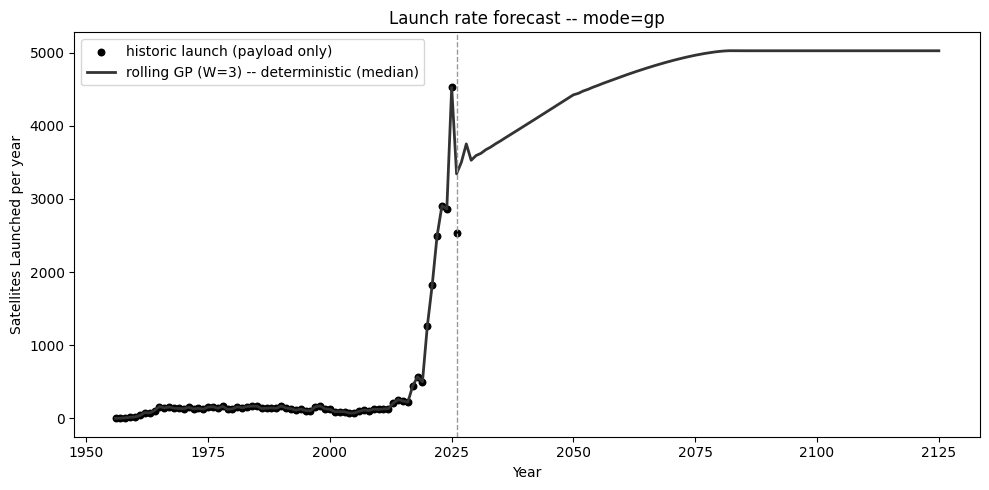}
  \caption{Time-varying launch rate forecast (satellites per year) obtained from
  a Gaussian Process trained on historical records 1956--2025, projected through
  2125.}
  \label{fig:gp_launch_rate}
\end{figure}

Launches enter the model as an exogenous, time-varying rate. 
We take the historical record of annual payload launches from 1957 to 2025 compiled by McDowell \cite{mcdowell2025_active_sats}, and extrapolate it forward with a Gaussian process (GP). 
The forecast proceeds one year at a time starting from the baseline year, 2025. At each step, the GP is fit to the three most recent years of data and predicts the following year's total. That prediction is added to the window before the next step. 
Therefore the projection track the recent acceleration in launch activity.
The annual launches through the end of the horizon as shown in Fig.~\ref{fig:gp_launch_rate}.
 
\begin{figure}[t]
  \centering
  \includegraphics[width=.5\textwidth]{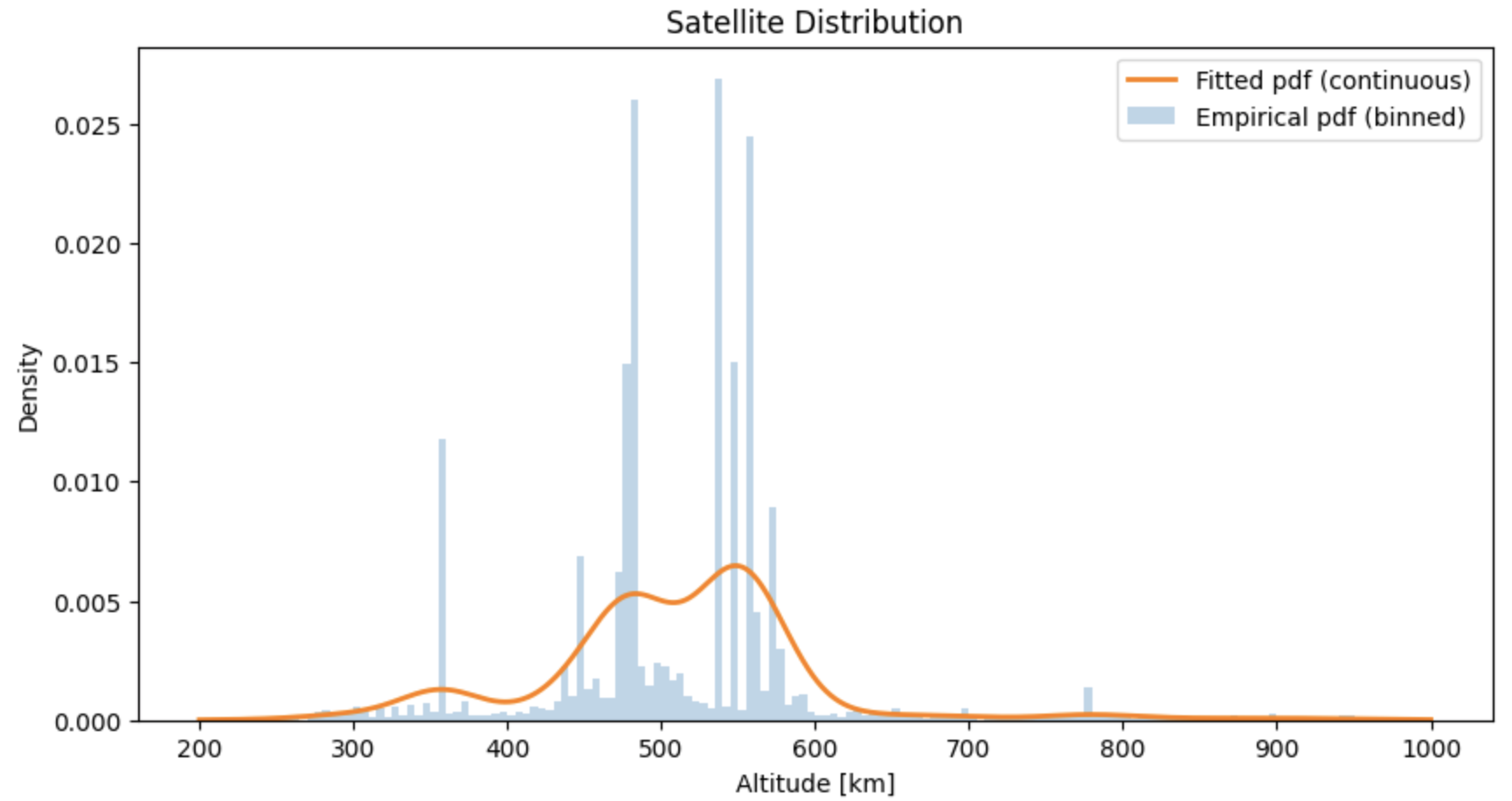}
  \caption{Empirical distribution of active satellites over LEO altitudes, used
  to spatially allocate the GP launch-rate forecast across shells.}
  \label{fig:sat_fit_alt}
\end{figure}

The annual total launch is distributed across altitude shells in proportion to the present-day active-satellite population illustrated in Fig.~\ref{fig:sat_fit_alt}, so that each shell receives a share of new launches matching where satellites are currently deployed. The launch input is thus data-driven and time-varying, and spatially allocated by the observed altitude distribution.

\subsection{Shell-Volume Scaling Design}
\label{sec:vscaling}

The central question is how the ODE--SDE discrepancy depends on per-shell
volume $V$. We address this by varying $V$ across three configurations while
holding the spatial density fixed: for each configuration the shell count $N$ is
chosen to partition the altitude band $450\,\mathrm{km}$--$800\,\mathrm{km}$ into shells of
the target volume, and each shell's initial population is scaled proportionally
to $V$ so that the per-volume density matches the baseline. The three
configurations are shown in Table~\ref{tab:vscaling}.
\begin{table}[t]
  \centering
  \caption{Shell-volume scaling configurations over the 450\,km--800\,km band.
  Population is scaled proportionally to $V$ at each configuration so that
  spatial density is conserved. Shell thickness decreases with altitude at
  fixed $V$; the range runs from the lowermost to the uppermost shell.}
  \label{tab:vscaling}
  \begin{tabular}{cccc}
  \hline
  Per-shell $V$ & $d$: lower $\rightarrow$ upper (km) & Average $d$ (km) & $N$ (shells) \\
  \hline
  $0.32\,\mathrm{M\,km}^3$ & $0.126 \rightarrow 0.040$ & $0.0635$ & $5510$ \\
  $3.2\,\mathrm{M\,km}^3$  & $1.26 \rightarrow 0.40$   & $0.635$  & $551$  \\
  $32\,\mathrm{M\,km}^3$   & $12.62 \rightarrow 4.01$  & $6.36$   & $55$   \\
  \hline
\end{tabular}
\end{table}

The $0.32\,\mathrm{M\,km}^3$ configuration, with shells of $126\,\mathrm{km}$ down to
$40\,\mathrm{m}$, corresponds to the narrow-shell regime characteristic of
modern megaconstellation spacing; the $32\,\mathrm{M\,km}^3$ configuration
matches the coarse shell width of established source--sink models such as MOCAT-3
\cite{d2023novel}. For each configuration, 50 independent SDE trajectories are
generated (50-run ensemble), and results are reported as ensemble means with
95\% pointwise confidence intervals.

Table~\ref{tab:params} lists the physical coefficients held fixed across
the sweep, following the parameterization of \cite{d2024carrying};
Table~\ref{tab:objprops} lists the representative mass, size, and area
assigned to each species, which set the impact parameter $\sigma_{ij}$
and enter the breakup model of Section~\ref{sec:mjp}.

We parameterize satellites and derelicts using the SpaceX Gen2 design, which represents a major share of current and planned deployments.
The smallest Gen2 variant filed with the FCC has a $2.8 \,\mathrm{m}\times 1.3\,\mathrm{m}$ bus and a mass of $303\,\mathrm{kg}$ \cite{spacex_fcc_letter_2022}.
To represent this rectangular bus in the collision model, we approximate it by a circle that encloses the bus. Its radius is half the diagonal of the rectangular bus, $r =\frac{1}{2}\sqrt{2.8^2+1.3^2}= 1.54\,\mathrm{m}$. We then use $A=\pi r^2$ as the representative cross-sectional area.
Debris retains the representative properties of \cite{d2024carrying}.

\begin{table}[t]
  \centering
  \caption{Physical parameters held fixed across the shell-volume sweep.}
  \label{tab:params}
  \begin{tabular}{lcc}
    \hline
    Parameter & Symbol & Value \\
    \hline
    Satellite operational lifetime & $\tau$   & $5\,\mathrm{years}$ \\
    Mean relative velocity         & $v$      & $10\,\mathrm{km/s}$ \\
    Satellite avoidance-failure fraction & $\alpha$   & $0.2$ \\
    Satellite--satellite avoidance-failure fraction & $\alpha_a$ & $0.01$ \\
    Disabling-to-lethal ratio      & $\delta$ & $10$ \\
    PMD success probability        & $P_M$    & $0.95$ \\
    Drag coefficient               & $c_D$    & $2.2$ \\
    Minimum fragment characteristic length & $L_C$ & $0.1\,\mathrm{m}$ \\
    \hline
  \end{tabular}
\end{table}

\begin{table}[t]
  \centering
  \caption{Representative physical properties by species. Satellite and
  derelict values follow the SpaceX Gen2 form factor filed with the FCC
  \cite{spacex_fcc_letter_2022}; derelicts share the mass and size of
  active satellites, since a derelict is a satellite that has failed to
  be disposed of successfully. Debris values follow \cite{d2024carrying}.
  Radii are the quantity entering the impact parameter
  $\sigma_{ij} = (r_i + r_j)^2$.}
  \label{tab:objprops}
  \begin{tabular}{lccc}
    \hline
    Species & Mass $M$ [kg] & Radius $r$ [m] & Area $A$ [m$^2$] \\
    \hline
    $S$ (satellite) & $303$  & $1.54$ & $7.48$ \\
    $D$ (derelict)  & $303$  & $1.54$ & $7.48$ \\
    $N$ (debris)    & $0.64$ & $0.09$ & $0.02$  \\
    \hline
  \end{tabular}
\end{table}

\subsection{Numerical Solvers and Convergence}

The ODE baseline is integrated with an adaptive embedded Runge--Kutta solver (Dormand--Prince). The SDE trajectories are integrated by the Euler--Maruyama scheme of Appendix~\ref{app:euler_maruyama} using a fixed step $\Delta t$. Convergence is verified by confirming that halving $\Delta t$ shifts ensemble means by less than one percent over the 100-year horizon.

The SDE approximation of the Markov jump in the narrow-shell regime is validated with a Gillespie-style discrete-event sampler (DES), which draws the exact jump times and jump sizes of Section~\ref{sec:mjp}. Because sampling every event individually is far more costly than stepping the diffusion, the check is run on a sub-band rather than the full band; it is reported in Section~\ref{sec:res_validation}.


\section{Results}
\label{sec:results}
 
The following results show the band resolution at three shell volumes, $V = 32$, $3.2$, and $0.32\,\mathrm{M\,km^3}$. Density is held fixed across configurations, so the three describe the same physical environment resolved at three scales, and the ODE baseline is common to all of them. Each configuration is a $50$-run ensemble.
 
\subsection{Validation against the Exact Jump Process}
\label{sec:res_validation}
 
The SDE is checked against the DES at the smallest shell volume, $V = 0.32\,\mathrm{M\,km^3}$.
Because the exact sampler is substantially more costly than Euler--Maruyama integration of the SDE approximation, the check is run at the uppermost $50\,\mathrm{km}$ of the band ($750\,\mathrm{km}$--$800\,\mathrm{km}$). Both ensembles place the debris population mean above the ODE trajectory---by an order of magnitude for the debris species at year $100$---and they agree within confidence intervals at year 50 and at year 100 as reported in Table~\ref{tab:desval}.
 
\begin{table}[t]
  \centering
  \caption{Discrete-event validation over the $750\,\mathrm{km}$--$800\,\mathrm{km}$
  band. Band-total populations from the ODE (ODE), the
  diffusion approximation (SDE, $100$ runs), and the exact discrete-event
  sampler (DES, $50$ runs).}
  \label{tab:desval}
  \begin{tabular}{llrcc}
    \hline
    Year & Species & ODE & SDE mean $[95\%\ \mathrm{CI}]$ & DES mean $[95\%\ \mathrm{CI}]$ \\
    \hline
    $50$  & $S$              & $259$     & $258$ $[251,\ 264]$             & $248$ $[233,\ 263]$ \\
          & $N$ ($0.64\,$kg) & $1{,}724$ & $2{,}895$ $[2{,}336,\ 3{,}501]$ & $2{,}671$ $[964,\ 5{,}953]$ \\
          & $N$ ($223\,$kg)  & $272$     & $366$ $[303,\ 433]$             & $280$ $[263,\ 301]$ \\
    \hline
    $100$ & $S$              & $263$     & $249$ $[239,\ 258]$             & $255$ $[231,\ 271]$ \\
          & $N$ ($0.64\,$kg) & $3{,}530$ & $36{,}533$ $[9{,}959,\ 86{,}000]$ & $26{,}969$ $[1{,}055,\ 78{,}647]$ \\
          & $N$ ($223\,$kg)  & $394$     & $639$ $[522,\ 764]$             & $430$ $[349,\ 532]$ \\
    \hline
  \end{tabular}
\end{table}
 
\subsection{Species-Dependent Response to Shell Volume}
\label{sec:res_species}
 
\begin{figure}[t]
  \centering
  \begin{subfigure}[t]{0.48\textwidth}
    \centering
    \includegraphics[width=\textwidth]{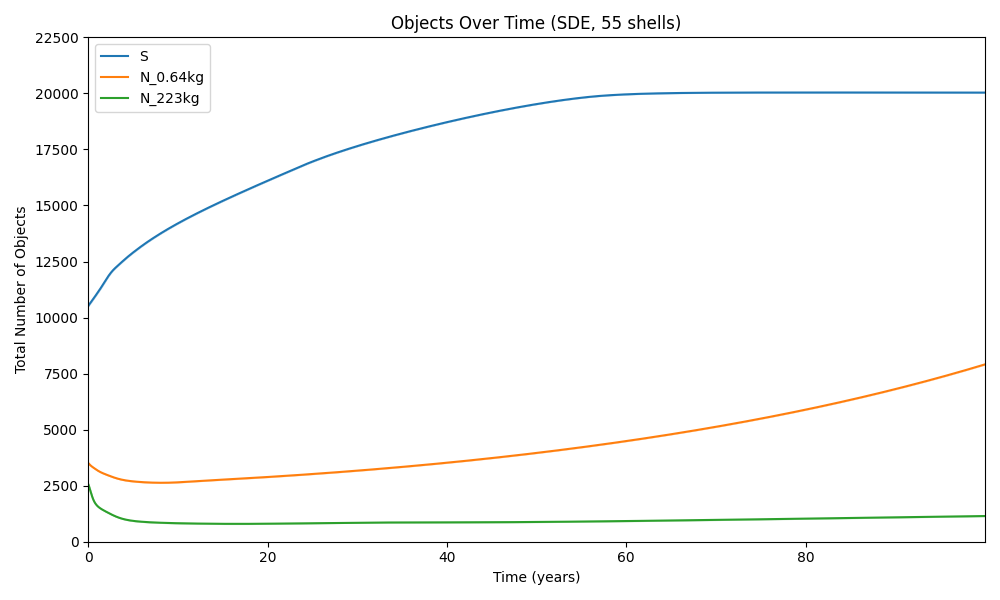}
    \caption{ODE}
    \label{fig:ens_ode}
  \end{subfigure}\hfill
  \begin{subfigure}[t]{0.48\textwidth}
    \centering
    \includegraphics[width=\textwidth]{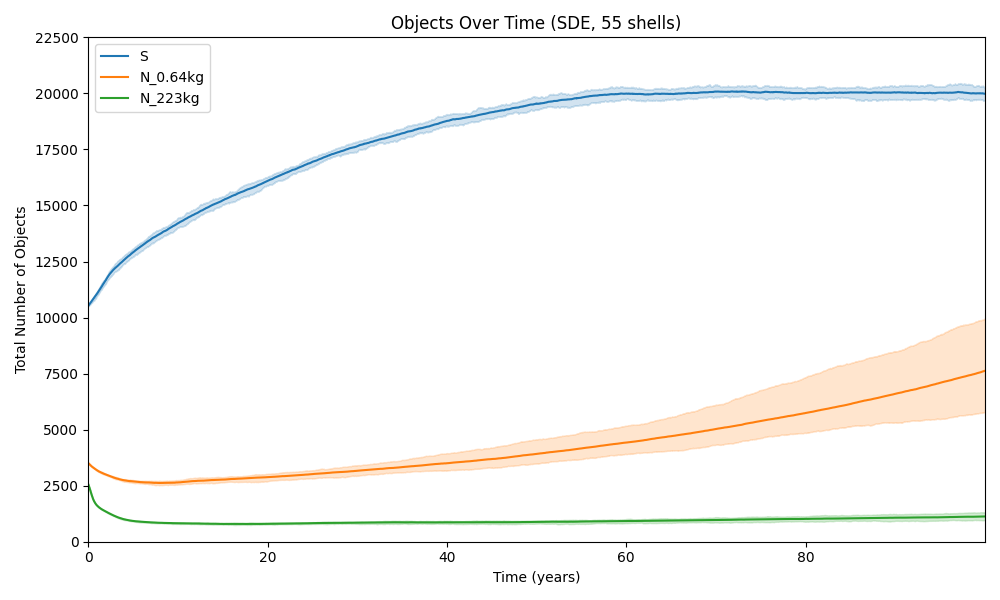}
    \caption{SDE, $V = 32\,\mathrm{M\,km^3}$}
    \label{fig:ens_32}
  \end{subfigure}
 
  \vspace{2pt}
 
  \begin{subfigure}[t]{0.48\textwidth}
    \centering
    \includegraphics[width=\textwidth]{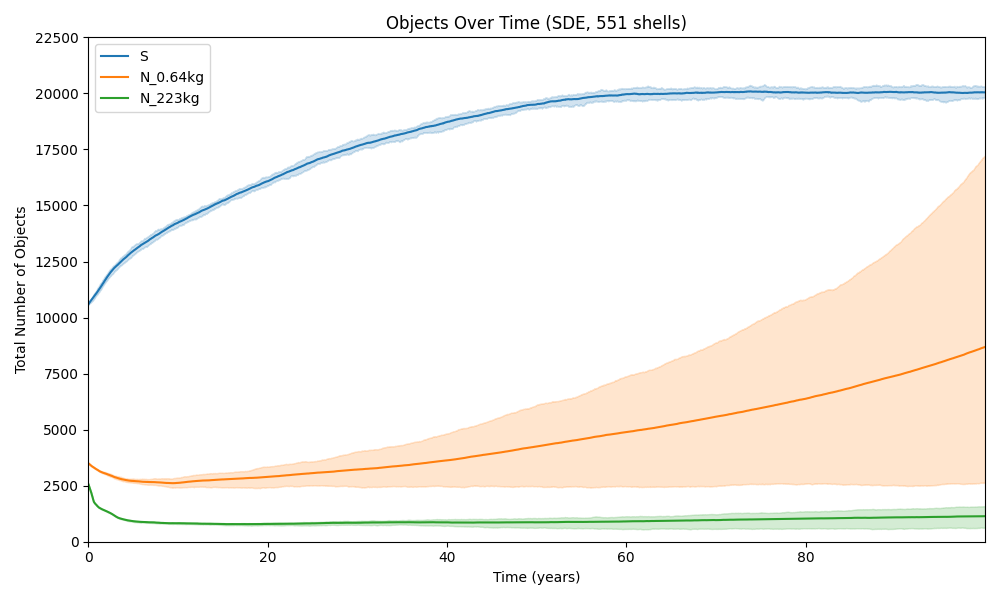}
    \caption{SDE, $V = 3.2\,\mathrm{M\,km^3}$}
    \label{fig:ens_3p2}
  \end{subfigure}\hfill
  \begin{subfigure}[t]{0.48\textwidth}
    \centering
    \includegraphics[width=\textwidth]{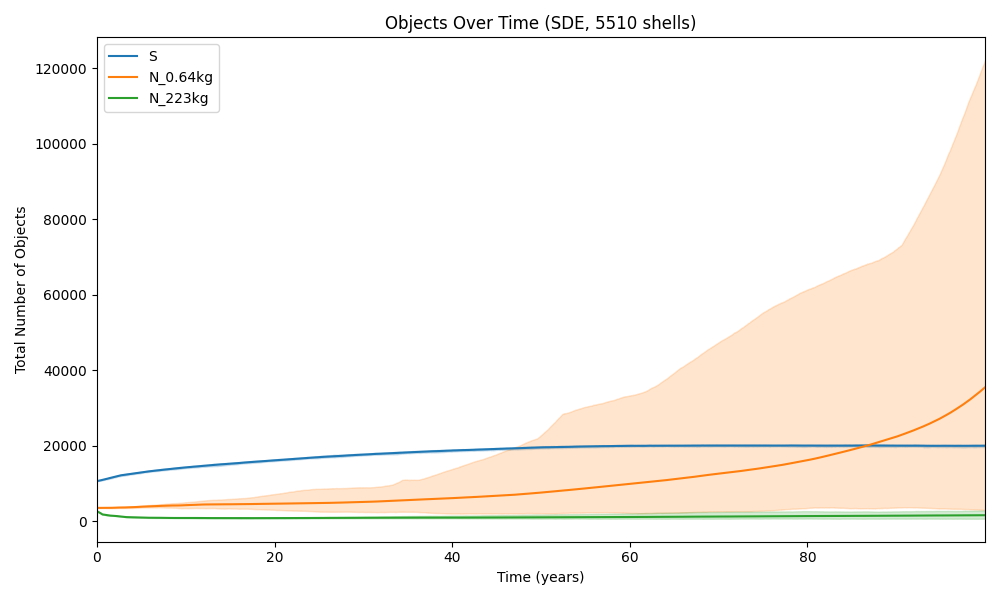}
    \caption{SDE, $V = 0.32\,\mathrm{M\,km^3}$ (y-axis differ in scale)}
    \label{fig:ens_0p32}
  \end{subfigure}
  \caption{Band-total populations over the $100$-year horizon for the
  deterministic solver (a) and the stochastic ensemble at the three
  shell volumes (b--d). Solid curves give ensemble means and shaded
  regions the $2.5$--$97.5$ percentile range over $50$ realizations.
  Panels (a)--(c) share a common vertical scale, on which the ensemble
  is nearly indistinguishable from the ODE; panel (d) is plotted on a
  scale roughly six times larger. The satellite and derelict
  populations are unchanged across all panels; the debris population
  broadens at $3.2\,\mathrm{M\,km^3}$ and both broadens and shifts
  upward at $0.32\,\mathrm{M\,km^3}$.}
  \label{fig:ensembles}
\end{figure}

Fig.~\ref{fig:ensembles} shows the ensemble trajectories over the
$450\,\mathrm{km}$--$800\,\mathrm{km}$ band at the three shell volumes.

The active satellite population rises from its initial value to roughly
$20{,}000$ objects by year $60$ and holds there, its ensemble band so
narrow as to be barely visible; the derelict species remains near
$10^3$ objects throughout. Neither responds drastically to the shell volume. The satellite population is governed by launch and disposal, which enter the model as processes external to the collision dynamics. The debris species is the one whose population responds to shell volume.
 
\subsection{Dispersion without Mean Departure}
\label{sec:res_dispersion}
 
At $V = 32\,\mathrm{M\,km^3}$ the ensemble reproduces the ODE trajectory. 
The mean of the debris population ends the horizon at $7{,}626$ objects against the deterministic $7{,}911$, a ratio of $0.96$, and the ensemble band remains a narrow envelope about the mean. 
At this shell volume the deterministic description is an adequate summary of the dynamics.
 
At $V = 3.2\,\mathrm{M\,km^3}$ the mean has moderately shifted, ending at $8{,}686$ objects, a ratio of $1.10$ that the ensemble cannot distinguish from unity. 
The confidence band, however, has changed noticeably: its upper edge now reaches roughly $17{,}000$ objects, about twice the mean, while its lower edge stays near the ODE path. 
The distribution has begun to stretch upward while its center has not yet followed.
Dispersion is the first thing to appear as shells narrow, and a study reporting only the mean at this shell volume would record no effect at all.
 
\subsection{Mean Departure at Narrow Shells}
\label{sec:res_departure}
 
At $V = 0.32\,\mathrm{M\,km^3}$ both the mean and the band has notably been displaced. The mean ends at $35{,}420$ objects, $4.5$ times the ODE value, and the upper
edge of the ensemble band reaches roughly $120{,}000$---six times the
full vertical scale of the two preceding configurations. The mean exceeds the ODE trajectory and the gap widens throughout.
 
The consequence is visible in the ordering of the species. Under the ODE
and at the two larger shell volumes alike, the debris population
remains well below the satellite population for the entire horizon. At
$0.32\,\mathrm{M\,km^3}$ the two cross at year $87$, after which the
mean debris count exceeds the mean number of active satellites in the
band.
 
\subsection{Satellite Losses}
\label{sec:res_attrition}
 
The satellite population is the same in all three configurations, but
the traffic that sustains it is not. Table~\ref{tab:attrition} compares
the realized collision events over the horizon between the ODE and the
smallest-volume SDE configuration. Satellites destroyed by collisions rise by a
factor of $2.1$ against derelicts and $7.7$ against debris, and total
satellite losses to collisions rise by a factor of $2.7$, while launches
over the same horizon are identical to within $0.1\%$. The
satellite--satellite rate, which depends on the satellite population
alone, is unchanged; the satellite--derelict and satellite--debris rates
rise because the populations they depend on have risen. Sustaining a
constellation in the narrow-shell environment costs more than twice as
many satellites, and the population level alone does not show it.
 
\begin{table}[t]
  \centering
  \caption{Realized collision events over the $100$-year horizon,
  ODE against $V = 0.32\,\mathrm{M\,km^3}$.
  Counts are ensemble means.}
  \label{tab:attrition}
  \begin{tabular}{lrrr}
    \hline
    Channel & ODE & $0.32\,\mathrm{M\,km^3}$ & Ratio \\
    \hline
    $S \times D$ & $325$ & $693$    & $2.1$ \\
    $S \times N$ & $45$  & $344$    & $7.7$ \\
    $S \times S$ & $12$  & $12$     & $1.0$ \\
    \hline
    Total        & $382$ & $1{,}048$ & $2.7$ \\
    \hline
  \end{tabular}
\end{table}
 
\subsection{Statistical Summary}
\label{sec:res_stats}
 
Table~\ref{tab:results} collects the comparison. At each shell volume we test whether the SDE ensemble mean at year $100$ differs from the ODE value, using a one-sample $t$-test against the null hypothesis that the $50$-run sample is drawn from a population with that mean; $p$ is the probability of a sample mean this far from the ODE value if the null hypothesis held. The two larger-volume configurations return no evidence against it ($p = 0.18$ and $p = 0.29$), while the smallest does ($p = 0.003$).
 
The fluctuations of Section~\ref{sec:sde_scaling} grow as shell volume
decreases, and the ensemble spreads accordingly. At
$32\,\mathrm{M\,km^3}$ no realization of debris population ends above twice the ODE debris population within the horizon; at $3.2\,\mathrm{M\,km^3}$ two of the $50$ do; at $0.32\,\mathrm{M\,km^3}$, $28$ realizations end above twice the ODE debris population and $8$ above ten times. The ODE, a single deterministic trajectory, represents none of this spread.
 
\begin{table}[t]
  \centering
  \caption{Debris population at year $100$ across shell volumes.
  Density is held fixed, so the ODE baseline
  ($7{,}911$ objects) is common to all three configurations. Means are
  reported with $95\%$ confidence intervals; medians are given for
  reference. $p$-values are from a one-sample $t$-test of the ensemble
  mean against the ODE value.}
  \label{tab:results}
  \begin{tabular}{lccccccc}
    \hline
    $V$ (M\,km$^3$) & $d$: lower $\rightarrow$ upper (km) & Shells & Mean & Ratio to ODE & $95\%$ CI & Median & $p$ \\
    \hline
    $32$   & $12.62 \rightarrow 4.01$ & $55$   & $7{,}626$  & $0.96$ & $\pm 406$      & $7{,}366$  & $0.18$ \\
    $3.2$  & $1.26 \rightarrow 0.40$  & $551$  & $8{,}686$  & $1.10$ & $\pm 1{,}404$  & $7{,}767$  & $0.29$ \\
    $0.32$ & $0.126 \rightarrow 0.040$ & $5510$ & $35{,}420$ & $4.48$ & $\pm 16{,}985$ & $18{,}376$ & $0.003$ \\
    \hline
  \end{tabular}
\end{table}
 
This randomness carries the mean of the debris itself away from the ODE at the smallest shell volume, reaching $4.48$ times the ODE debris population, while the two larger-volume configurations hold near the ODE results. Three realizations at the smallest shell volume overflow numerical bounds before year $100$; none does at either larger shell volume.

\section{Discussion}
\label{sec:discussion}

Established models discretize altitude into bins of tens of
kilometers. Our largest shell volume configuration features shell thicknesses of $12.6$--$4.0\,\mathrm{km}$, which is at the fine end of that range. The effect that appears at
$0.32\,\mathrm{M\,km^3}$ is not apparent at larger-volume configurations, since averaging over a large volume suppresses the shell-scale randomness that drives it. As operators and regulators move toward the kilometer and sub-kilometer separations, the scale at which capacity is debated approaches the scale at which the deterministic description begins to lose accuracy.
 
\subsection{Mechanism of the Stochasticity Gap}
\label{sec:origin_gap}

The ODE and SDE description differs in how averages are evaluated. 
Taking an expectation of \eqref{eq:Z}, the diffusion term has zero mean, and under standard integrability conditions that allow expectation and time integration to be exchanged, the ensemble mean of the SDE and the ODE satisfy
\begin{align}
  \mathbb{E}\bigl[Z_{V,Q}(t)\bigr]
    &= x_Q(0) + \int_0^t \mathbb{E}\bigl[F_Q(Z_V(u))\bigr]\, du,
  \label{eq:mean_sde} \\[2pt]
  x_Q(t)
    &= x_Q(0) + \int_0^t F_Q\bigl(x(u)\bigr)\, du,
  \label{eq:mean_ode}
\end{align}
respectively. Equations \eqref{eq:mean_sde} and \eqref{eq:mean_ode} differ in the argument involving $F_Q$. 
In \eqref{eq:mean_ode}, the ODE evaluates the drift at the deterministic population density $x(u)$, while in \eqref{eq:mean_sde} the SDE averages over stochastic $Z_V$.
Their difference, $\mathbb{E}[F_Q(Z_V)] - F_Q(\mathbb{E}[Z_V])$, generates the stochasticity gap.

The stochasticity gap is confined to the collision channels. Indeed, the collision intensities are products of two densities, $\beta_c(x) = k_c\, x_{i(c)} x_{j(c)}$ with $k_c = \gamma_c\, \sigma_{i(c)j(c)}\, v$, where $\gamma_c \in \{1, \alpha, \delta, \alpha_a\}$ is the coefficient related to channel $c$, and $i(c)$ and $j(c)$ are the species colliding in channel $c$.
For a product
$\mathbb{E}[Z_{V,i} Z_{V,j}]
 = \mathbb{E}[Z_{V,i}]\,\mathbb{E}[Z_{V,j}]
 + \mathrm{Cov}(Z_{V,i}, Z_{V,j})$. Weighting each channel by its jump size and summing over the collision channels $\mathcal{C}$,
\begin{equation}
  \mathbb{E}\bigl[F_Q(Z_V)\bigr] - F_Q\bigl(\mathbb{E}[Z_V]\bigr)
    = \sum_{c \in \mathcal{C}} \ell_{c}\, k_c\,
      \mathrm{Cov}\bigl(Z_{V,i(c)},\, Z_{V,j(c)}\bigr).
  \label{eq:jensen_gap}
\end{equation}

The covariance term depends on the shell volume. Section~\ref{sec:sde_scaling} establishes the fluctuation amplitude of each density as $O(V^{-1/2})$; 
since that amplitude is a standard deviation, the corresponding variance, and likewise the covariance between two densities, is $O(V^{-1})$. The right-hand side of \eqref{eq:jensen_gap} therefore grows as shells narrow, and tends to zero as $V \to \infty$, consistent with the limit of Section~\ref{sec:mjp_limit}.

Debris production makes the difference in \eqref{eq:jensen_gap} cumulative. 
The collision rate involving debris rises with the debris density. 
The stochastic variation in \eqref{eq:jensen_gap} adds a positive increment to the collision rate. 
The increment produces extra debris, which is carried forward and raises the collision rate again. 
Each infinitesimal contribution of debris in the collision rate accumulates and results in larger debris and derelict population. 
The departure from the ODE compounds over the horizon, from $0.96$ of the ODE value at $32\,\mathrm{M\,km^3}$ to $4.48$ at $0.32\,\mathrm{M\,km^3}$ (Table~\ref{tab:results}). 
Table~\ref{tab:attrition} shows the effect of this mechanism channel by channel. 
The satellite-satellite collision rate, which involves only the satellite density, is unchanged, while the rates involving a derelict or debris density rise with them.

\subsection{Implications for Satellite Losses}

Table~\ref{tab:attrition} quantifies the cost of the rising satellite--derelict and satellite--debris collision rates. 
Over the horizon, satellites destroyed by collision rise from $382$ under the ODE to $1{,}048$ in the fine-shell environment, a $2.7\times$ increase.
This is an operational and economic cost borne by operators, and the deterministic model understates this.

Following the analysis in \eqref{eq:jensen_gap}, the ODE evaluates the
collision rate at the mean population and therefore underestimates the
collision count.
Section~\ref{sec:origin_gap} showed that the debris is the fastest growing population with high variance.
The underestimated collision count of the ODE poses a significant threat to the scenario planning in every shell.

\subsection{Implications for Launch and Disposal Policy}

In previous works, launch planning and PMD success rate targets have been assessed against the deterministic prediction of long-term stability.
Since the SDE predicts a larger debris and derelict population than the ODE at smaller shell volumes, a policy calibrated on the ODE will have to be adjusted. 
Specifically, we explore how launch cadence and PMD success rate would have to be adjusted for the SDE outcomes to meet the ODE debris and derelict population levels. 
We set the baseline policy as the settings used throughout the preceding sections, including 
the data-driven launch schedule of Section~\ref{sec:launch_rate} and a PMD success rate of $P_M = 0.95$ as in Table ~\ref{tab:params}. 
Launch cadence is varied through a multiplier $m$ applied to the baseline schedule, so that $m = 1$ reproduces the baseline and $m < 1$
scales launch rate of every shell down in proportion. PMD success rate is varied directly through $P_M$, with larger values raising the fraction of satellites successfully removed at end of life, so that fewer enter the derelict population.

We then examine how much $m$ must fall, or how much $P_M$ must rise, for the SDE ensemble mean of the combined derelict and debris population ($D + N$) at year $100$ to return to the value the ODE predicts for that same population under the baseline policy. 
We report the intermediate ($V = 3.2\,\mathrm{M\,km^3}$) and smallest ($V = 0.32\,\mathrm{M\,km^3}$) configurations, and state the results qualitatively.

\begin{figure}[t]
  \centering
  \includegraphics[width=0.48\textwidth]{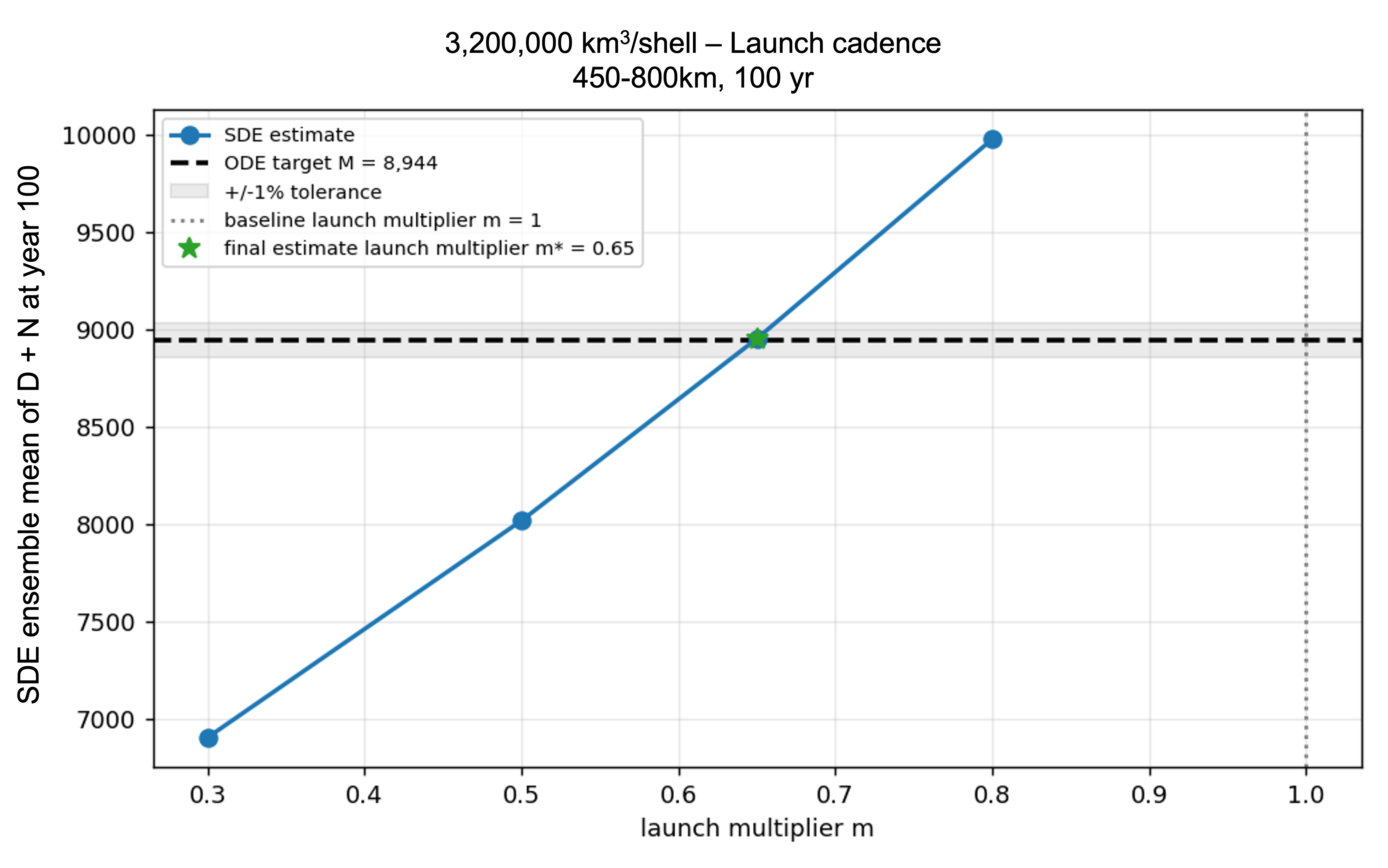}
  \includegraphics[width=0.48\textwidth]{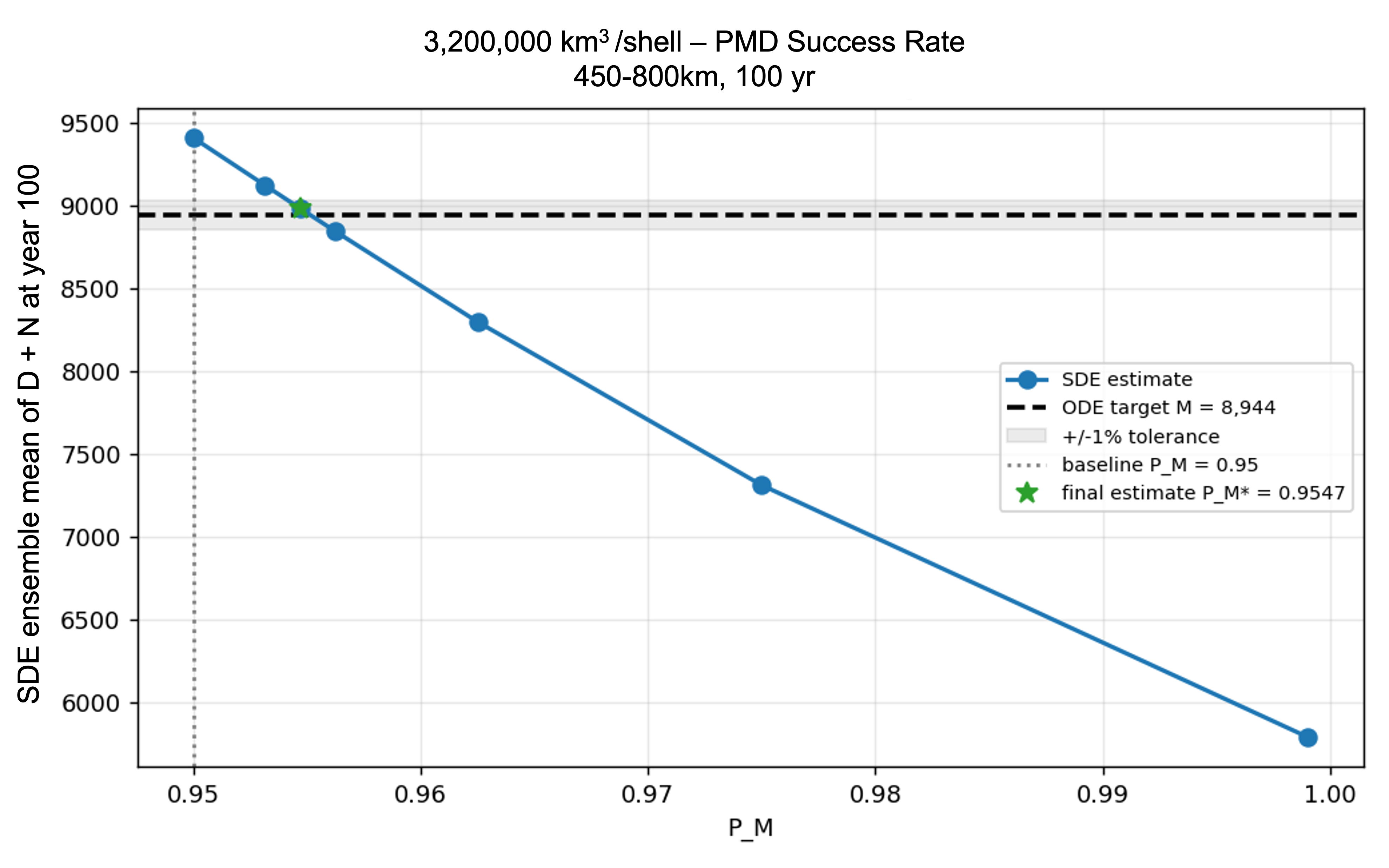}
  \caption{Compensating policy adjustment at $V =
  3.2\,\mathrm{M\,km^3}$. The SDE ensemble mean of the combined
  derelict and debris population at year $100$ is shown against launch
  multiplier (left) and disposal success rate (right); the dashed line
  is the ODE prediction under the baseline policy. Both axes
  cross it.}
  \label{fig:bisection_3p2}
\end{figure}

\paragraph{Launch cadence.} At $V =
3.2\,\mathrm{M\,km^3}$ both launch cadence and PMD success rate adjustments help reach the target. Lowering the launch rate to about two-thirds of the baseline level, the ODE debris and derelict level can be recovered (Fig.~\ref{fig:bisection_3p2}).

\begin{figure}[t]
  \centering
  \includegraphics[width=0.48\textwidth]{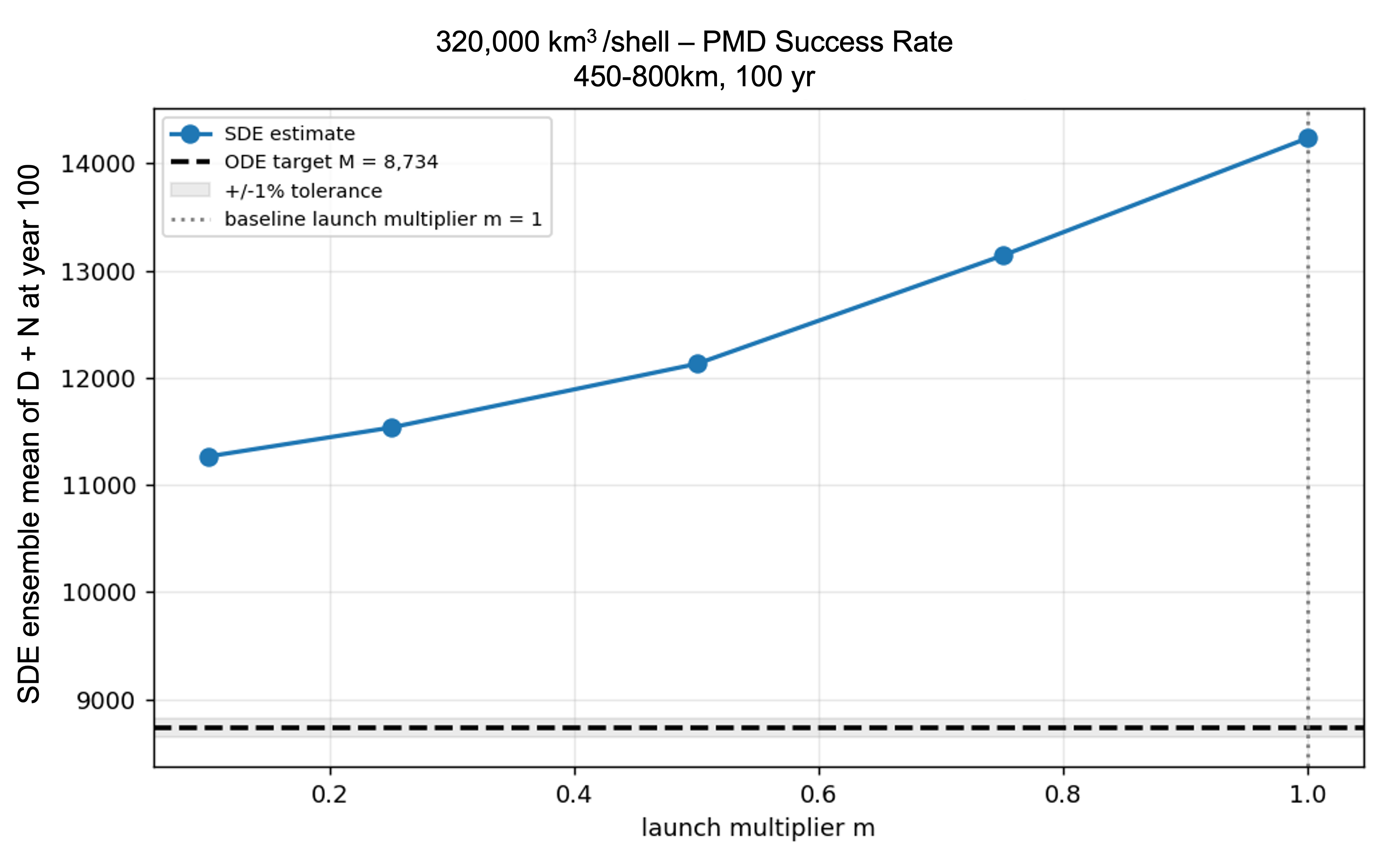}
  \includegraphics[width=0.48\textwidth]{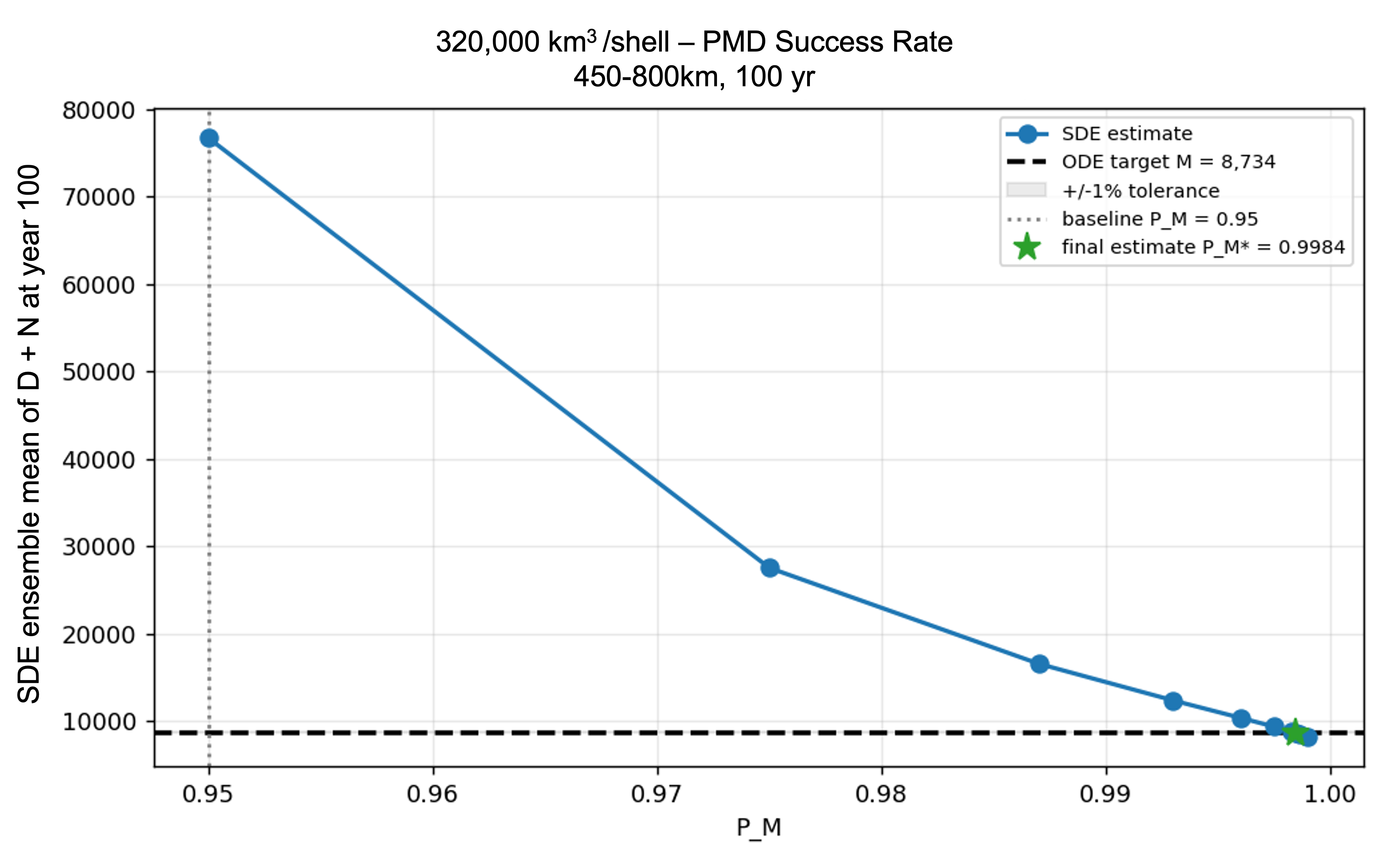}
  \caption{Compensating policy adjustment at $V =
  0.32\,\mathrm{M\,km^3}$. The launch axis (left) does not reach the
  ODE target anywhere in the range searched and flattens as
  cadence is reduced; the disposal axis (right) reaches it only as
  reliability approaches unity.}
  \label{fig:bisection_0p32}
\end{figure}

At $V = 0.32\,\mathrm{M\,km^3}$, the ODE debris and derelict level cannot be reached by lowering the launch cadence.
Reducing launch cadence lowers the year-$100$ population.
However, the reduction in the debris population decreases. 
Cutting launches to a quarter of the baseline recovers most of the available reduction, and cutting them further to a tenth moves the mean of the debris and derelict by only a further two percent (Fig.~\ref{fig:bisection_0p32}). 
Over the entire range searched, no launch cadence brings the SDE ensemble mean back to the ODE value.

\paragraph{PMD success rate.} The PMD requirements escalates sharply from $V = 3.2\,\mathrm{M\,km^3}$ to $V = 0.32\,\mathrm{M\,km^3}$.
At $V = 3.2\,\mathrm{M\,km^3}$ a fraction of a percentage point suffices. 
At $V = 0.32\,\mathrm{M\,km^3}$ the PMD success rate must approach perfection. 
When expressed as a failure rate, the share of satellites failing disposal must fall from one in twenty to roughly one in several hundred. This is a reduction of more than an order of magnitude, and a standard far beyond current compliance.

At $V = 3.2\,\mathrm{M\,km^3}$, reducing launch cadence and raising PMD success rate substitute for one another. 
The comparison between launch cadence and PMD success rate is informative, since a sub-percentage-point improvement in disposal reliability does the work of a one-third reduction in launch traffic. 
At $V = 0.32\,\mathrm{M\,km^3}$ the equivalent fails. 
Launch cadence ceases to be an instrument at all. PMD success rate remains, but the standard it must meet approaches perfection. 
A policy discussion that treats launch pacing and disposal compliance as alternative routes to the same environmental outcome is making an assumption that holds at large shell volumes and fails at the volumes toward which orbital traffic is moving.

The results in this subsection are qualitative.
 The launch cadence reduction and PMD success rate are provided as point estimates, but the variance around the estimates is large at the examined shell volumes. 
The implications are intended to show the direction of each adjustment and its rough magnitude. Resolving the policy implications into quantitative thresholds would require variance reduction in the ensemble estimates, which we leave to future work.

\subsection{Limitations}

The diffusion approximation was validated against the discrete-event system (DES) at the smallest shell volume under consideration. 
The validation was carried out over the uppermost 50km of the simulated altitudes, from 750km to 800km, which was chosen because the exact sampler is costly to run. 
A check across the remaining altitudes would place the SDE approximation on firmer ground. 

Furthermore, in the simulation on 450km-800km altitude range, three of 50 SDE realizations overflowed numerical bounds before the end of the horizon and are excluded from the reported statistics. 
The excluded realizations feature the fastest-growing debris species. Consequently, the reported mean of debris and derelict populations in the SDE is likely a slight underestimate.


\section{Conclusion}
\label{sec:conclusion}
Established source--sink models describe the LEO environment by
propagating shell-averaged populations as smooth deterministic flows.
This description is valid under broad-shell control and when the events occurring within are numerous.
This work examined the LEO environment at small-volume shells into which orbital traffic is actively being organized.
We formalized the collision, drag, launch, and PMD effects in a multi-shell, multi-species environment as a MJP. 
At the large-shell volume limit, the fluid-scaled MJP recovers the deterministic source-sink equations of established models.
At smaller shell volumes the diffusion-scaled MJP yields a diffusion approximation that retains the finite-volume fluctuations the deterministic description fails to represent. 
 
We find that the deterministic and stochastic descriptions agree at the volumes current modeling approaches use but separate at the volumes toward which constellation design is moving. 
At intermediate volumes a moderate departure of the ASO population from the ODE is present in individual realizations, while the mean population of each species is not statistically distinguishable from the ODE value. 
At the narrowest volumes the mean of the debris population shifts significantly, reaching $4.48$ times the ODE value. 
Many realizations end at several times the debris population of the ODE solution at year $100$. 
The effect enters through the collision terms, producing a large number of debris that feeds back into the system.
 
The satellite population comparison demonstrates what can be missed in the deterministic description. The active satellite count is near identical under both ODE and SDE descriptions. Yet the number of satellites destroyed by collision over the horizon is $2.7$ times larger in the finest shell volume, driven by the collision rate involving debris. Randomness at the shell scale is registered not in the size of the active population but in the operational and economic cost of the losses it sustains, which is underestimated by the deterministic ODE.
 
These findings bear on decisions being made at the shell scale. 
Capacity limits, collision-risk thresholds, and disposal-compliance targets are increasingly argued at vertical separations of a few kilometers or less, precisely where our results place the boundary of the deterministic description. 
A shell that an ODE model reports as stable may carry a non-negligible probability of runaway growth once finer-scale approximations are taken into account. 
Because the stochastic model shares its parameterization with the ODE, it should be viewed as a check on conclusions already drawn rather than as a replacement for the models that produced them.
 
A direction for future work is to connect the SDE framework to orbital slotting, informed by the quantitative analysis of launch and disposal adjustment policies. 
Slotting schemes raise the capacity of a band by placing satellites in ever thinner shells~\cite{lifson2023space}.
Our results show that smaller-volume shells raise the risk of collision within each of them.
This trade-off motivates slotting and launch-pacing strategies that choose the shell width and launch profiles together, so that the geometric capacity of a shell and the long-term stability of the LEO environment are jointly designed.

\bibliography{references}



\appendix
\section{Channel Enumeration and Transition Intensities}
\label{app:channels}
 

\subsection{Collision Intensities from the Binomial Argument}
 
The collision intensities of Table~\ref{tab:allchannels} follow from a
binomial argument. Treating each of the $D$ derelicts as an independent
candidate for a collision over a short interval $h$, with per-object
probability $p = \sigma_{ij}\, v\, x_j\, h$ of encountering field
species $j$, the probability of exactly one derelict-removing event is
\begin{align}
  P\bigl(\hat{X}_{V,D}(t+h) = D-1 \mid \hat{X}_V(t) = (S,D,N)\bigr)
  &= \binom{D}{1}\Bigl(\sigma_{DN} v\, \tfrac{N}{V} h\Bigr)^{1}
      \Bigl(1 - \sigma_{DN} v\, \tfrac{N}{V} h\Bigr)^{D-1} \notag \\
  &\quad + \binom{D}{1}\Bigl(\alpha\,\sigma_{DS} v\, \tfrac{S}{V} h\Bigr)^{1}
      \Bigl(1 - \alpha\,\sigma_{DS} v\, \tfrac{S}{V} h\Bigr)^{D-1} \notag \\
  &\quad + \Bigl(\tfrac{\omega}{d} D\, h + o(h)\Bigr) \notag \\
  &= V h\Bigl(\sigma_{DN} v\, \tfrac{N}{V}\tfrac{D}{V}
      + \alpha\,\sigma_{DS} v\, \tfrac{S}{V}\tfrac{D}{V}\Bigr)
      + \tfrac{\omega}{d} D\, h + o(h),
  \label{eq:D_minus1_appendix}
\end{align}
where the second equality keeps only the $O(h)$ terms: each binomial
factor contributes its mean $D \cdot (\sigma_{ij} v\, x_j h)$ to
leading order, and the product $D \cdot x_j$ reorganizes into
$V \cdot x_i x_j$ since $x_i = D/V$. The gain channels,
\begin{align}
  P\bigl(\hat{X}_{V,D}(t+h) = D+1 \mid \hat{X}_V(t) = (S,D,N)\bigr)
  &= V h\Bigl(\delta\,\sigma_{SN} v\, \tfrac{N}{V}\tfrac{S}{V}
      + \delta\,\sigma_{DS} v\, \tfrac{D}{V}\tfrac{S}{V}
      + (1-P_M)\tfrac{S}{V}\tfrac{1}{\tau}\Bigr)
      + \tfrac{\omega_+}{d_+} D_+\, h + o(h),
  \label{eq:D_plus1}
\end{align}
and the paired loss
\begin{align}
  P\bigl(\hat{X}_{V,D}(t+h) = D-2 \mid \hat{X}_V(t) = (S,D,N)\bigr)
  &= V h\,\sigma_{DD} v\, \tfrac{D}{V}\tfrac{D}{V} + o(h),
  \label{eq:D_minus2}
\end{align}
are constructed identically. All remaining channels of
Table~\ref{tab:allchannels} follow the same pattern.
 
\subsection{Law of Large Numbers}
 
With $F(x) = \sum_c \ell_c \beta_c(x)$ and $x(t)$ solving
$x(t) = x(0) + \int_0^t F(x(s))\, ds$, the Kurtz--Ethier theorem
\cite{ethier2009markov,Draief_Massoulie_2009} gives
\begin{equation}
  \lim_{V \to \infty}\ \sup_{0 \le t \le T}
    \left\| \tfrac{1}{V}\hat{X}_V(t) - x(t) \right\| = 0
  \quad \text{a.s.}
  \label{eq:kurtz}
\end{equation}

 
\section{Formal Construction of the Diffusion Approximation}
\label{app:sde}
 
\subsection{Centered Jump Representation}
 
Write the density process $X_{V,Q}(t) = \hat{X}_{V,Q}(t)/V$ in terms
of the channels $c$ for species $Q$, each with jump size $\ell_c$ and
density-dependent intensity $\beta_{c,Q}(x)$, driven by an
independent unit-rate Poisson process $Y_{c,Q}$, and let
$F_Q(x) = \sum_{c} \ell_c\, \beta_{c,Q}(x)$ denote the net drift as
in Section~\ref{sec:mjp}. Centering each Poisson process about its
mean, $\tilde{Y}_{c,Q}(\theta) = Y_{c,Q}(\theta) - \theta$, separates
the density into a fluctuation term and a drift integral,
\begin{equation}
  X_{V,Q}(t) = X_{V,Q}(0)
    + \frac{1}{V}\sum_{c} \tilde{Y}_{c,Q}\!\left(V \int_0^t
      \beta_{c,Q}(X_V(u))\, du\right)
    + \int_0^t F_Q(X_V(u))\, du.
  \label{eq:sde_centered}
\end{equation}
As $V \to \infty$ the centered term vanishes at rate $V^{-1/2}$ and
$X_{V,Q}$ converges to the ODE trajectory of
Section~\ref{sec:mjp}, the content of the law of large numbers.
 
\subsection{Diffusion Limit}
 
By the functional central limit theorem for density-dependent Markov
chains \cite{ethier2009markov}, the rescaled process
$V^{-1/2}\tilde{Y}_{c,Q}(V\theta)$ converges in distribution to a
Brownian motion $W_{c,Q}$ whose variance accumulates at rate
$\beta_{c,Q}(X(u))$ along the trajectory. The time-change identity
\begin{equation}
  W_{c,Q}\!\left(\int_0^t \beta_{c,Q}(X(u))\, du\right)
  = \int_0^t \beta_{c,Q}^{1/2}(X(u))\, dW_{c,Q}(u),
  \label{eq:tcid}
\end{equation}
which holds in distribution by equality of quadratic variation,
converts the time-changed Brownian motion into It\^{o} form and
yields the closed approximation \eqref{eq:Z} of the main text.
 
\subsection{Full System with Drag}
 
Carrying the drag clocks through the same centering and $\sqrt{V}$
scaling adds a drag drift and a matching diffusion term to
\eqref{eq:Z},
\begin{equation}
  \begin{aligned}
    Z_{V,Q}(t) = X_{V,Q}(0)
    &+ \int_0^t \Bigl(F_Q(Z_V(u)) + \beta_{\mathrm{drag},Q}^{+}(Z_{V,Q}(u))
        - \beta_{\mathrm{drag},Q}^{-}(Z_{V,Q}(u))\Bigr) du \\
    &+ \frac{1}{\sqrt{V}} \sum_{c} \ell_c \int_0^t
        \beta_{c,Q}^{1/2}(Z_V(u))\, dW_{c,Q}(u) \\
    &+ \frac{1}{\sqrt{V}} \int_0^t
        \Bigl(\bigl(\beta_{\mathrm{drag},Q}^{+}(Z_{V,Q}(u))\bigr)^{1/2}
        - \bigl(\beta_{\mathrm{drag},Q}^{-}(Z_{V,Q}(u))\bigr)^{1/2}\Bigr)
        dW_{\mathrm{drag},Q}(u),
  \end{aligned}
  \label{eq:Z_drag}
\end{equation}
with $\beta_{\mathrm{drag},Q}^{-} = \tfrac{\omega}{d}\,x$ and
$\beta_{\mathrm{drag},Q}^{+} = \tfrac{\omega_+}{d_+}\,x_+$ as in the main text.

\subsection{Numerical Discretization}
\label{app:euler_maruyama}
 
Trajectories of \eqref{eq:Z_drag} are simulated by the Euler--Maruyama scheme, the standard first-order method for numerically integrating stochastic differential equations~\cite{higham2001algorithmic}. Writing the drift and diffusion compactly as $F(\cdot)$ and $G(\cdot)$, with $G(\cdot) \propto \beta(\cdot)^{1/2}$ collecting the square-root coefficients of all collision, drag, and launch channels, the update over a uniform step $\Delta t$ is
\begin{equation}
  Z_V^{n+1} = Z_V^{n} + F(Z_V^{n})\, \Delta t + G(Z_V^{n})\, \Delta W^n,
  \label{eq:euler_maruyama}
\end{equation}
where $\Delta W^n \sim \mathcal{N}(0, \Delta t\, I)$ is a vector of independent Gaussian increments, one per channel, with $\mathcal{N}(\mu, \Sigma)$ denoting the Gaussian distribution of mean $\mu$ and covariance $\Sigma$. Each increment has mean zero and variance $\Delta t$, the $\sqrt{\Delta t}$ scale of the increment being the discrete signature of the diffusive fluctuations.

\end{document}